\documentclass[twocolumn,superscriptaddress,floatfix]{revtex4-1}
\usepackage{graphicx}
\usepackage{dcolumn}
\usepackage{bm}
\usepackage{multirow}
\usepackage{amssymb}
\usepackage{amsmath}
\usepackage{graphicx}
\usepackage{enumerate}
\usepackage[colorlinks,citecolor=blue,linkcolor=red,urlcolor=blue]{hyperref}

\begin{document}
\title{Equilibration of Topological Defects Near the Deconfined Quantum Multicritical Point}
\author{Yu-Rong Shu}
\affiliation{School of Physics and Materials Science, Guangzhou University, Guangzhou 510006, China}
\author{Shao-Kai Jian}
\affiliation{Department of Physics and Engineering Physics, Tulane University, New Orleans, Louisiana, 70118, USA}
\author{Anders W. Sandvik}
\email{sandvik@bu.edu}
\affiliation{Department of Physics, Boston University, Boston, Massachusetts 02215, USA}
\author{Shuai Yin}
\email{yinsh6@mail.sysu.edu.cn}
\affiliation{School of Physics, Sun Yat-Sen University, Guangzhou 510275, China}
\affiliation{Guangdong Provincial Key Laboratory of Magnetoelectric Physics and Devices, Sun Yat-Sen University, Guangzhou 510275, China}
\date{\today}

\begin{abstract}
Deconfined quantum criticality (DQC) arises from fractionalization of quasi-particles and leads to fascinating behaviors beyond the Landau-Ginzburg-Wilson description of phase transitions. Here, we study the critical dynamics when driving a two-dimensional quantum magnet through a weakly first-order transition point near a putative deconfined multicritical point separating antiferromagnetic and spontaneously dimerized ground states. Numerical simulations show that the conventional Kibble-Zurek scaling (KZS) mechanism is inadequate for describing the annealing process. We introduce the concept of dual asymmetric KZS, where both a pseudocritical relaxation time and the deconfinement time enter and the scaling also depends on the driving direction according to a duality principle connecting the topological defects in the two phases. These defects require a much longer time scale for equilibration than the amplitude of the order parameter. Beyond advancing the DQC scenario, our scaling approach provides a new window into out-of-equilibrium criticality with multiple length and time scales. 
\end{abstract}

\maketitle

\noindent{\bf INTRODUCTION}\\
The concept of deconfined quantum criticality (DQC) was introduced \cite{Senthil2004,Levin2004prb,Senthil2006prb} as a paradigm beyond that
of Landau-Ginzburg-Wilson (LGW) for certain continuous phase transitions between ordered ground states with unrelated broken symmetries.
Though evidence for critical points with fractionalized excitations and emergent gauge fields has been found in simulations of lattice models
\cite{Sandvik2007prl,Melko2008prl,Lou2009prb,Chen2013prl,Harada2013prb,Nahum2015prx,Nahum2015prl,Shao2016,Ma2018prb}, the exact nature of the
DQC phenomenon is still under intensive scrutiny \cite{Wang2017prx,Ihrig2019prb,Zhao2020prl,Ma20,Nahum20,He21,Wang21,Lu21,Yuan23,Christos2023pnas}.

In the paradigmatic example of DQC between the antiferromagnetic (AFM) and spontaneously dimerized valence-bond-solid (VBS) phases of
spin-isotropic $S=1/2$ magnets on the two-dimensional ($2$D) square lattice, the quantum numbers carried by topological defects in one phase
correspond to the order parameter in the other phase. Specifically, the theory posits that fractionalized spin excitations (spinons) in the
VBS phase and space-time hedgehog singularities (monopoles) in the AFM phase deconfine upon approaching the critical point and proliferate when
crossing the phase transition, thereby inducing the complementary order parameter \cite{Senthil2004,Levin2004prb}. Quantum interference between Berry phases of monopoles makes
their fugacity irrelevant at the critical point \cite{Senthil2004,Levin2004prb}. The associated emergent U$(1)$ symmetry of the near-critical VBS
ground state (where quadrupled monopoles are dangerously irrelevant) was confirmed numerically \cite{Sandvik2007prl,Lou2009prb} and a higher SO$(5)$ symmetry
at the critical point in one variant of the theory \cite{Senthil2006prb} has also been detected \cite{Nahum2015prl,Takahashi2020prr}.

The DQC theory was challenged by the contradiction between the values of critical exponents estimated in the $JQ$ model \cite{Sandvik2007prl}, which harbors a direct AFM-VBS phase transition and was regarded as a typical lattice model realizing the DQC, and their bounds given by
numerical bootstrap calculations for an SO$(5)$-symmetric conformal field theory (CFT) under the assumption of a single relevant
direction~\cite{Nakayama2016prl}. Many other studies also point to an ultimately discontinuous phase
transition~\cite{Jiang2008jst,Chen2013prl,Li2022jhep,DEmidio2023sp,Chen2024prl,Takahashi2024}. In particular, the entanglement entropy for a corner-less
bipartition at the transition point of the $JQ$ model shows the signature of four Goldstone modes~\cite{Deng2024prl}, indicating the
coexistence of the AFM and VBS orders, while the corner contributions for a tilted bipartition exhibits a critical form consistent
with an SO(5) CFT \cite{DEmidio2024prl}. These and other results \cite{Takahashi2024} strongly suggest that, while the AFM-VBS transition
in the $JQ$ model is a weakly first-order phase transition (FOPT), it is extremely close to an SO$(5)$ multicritical point and
exhibits critical scaling on distances up to hundred or more lattice spacings.

\begin{figure*}[t]
\centering
\includegraphics[width=\linewidth,clip]{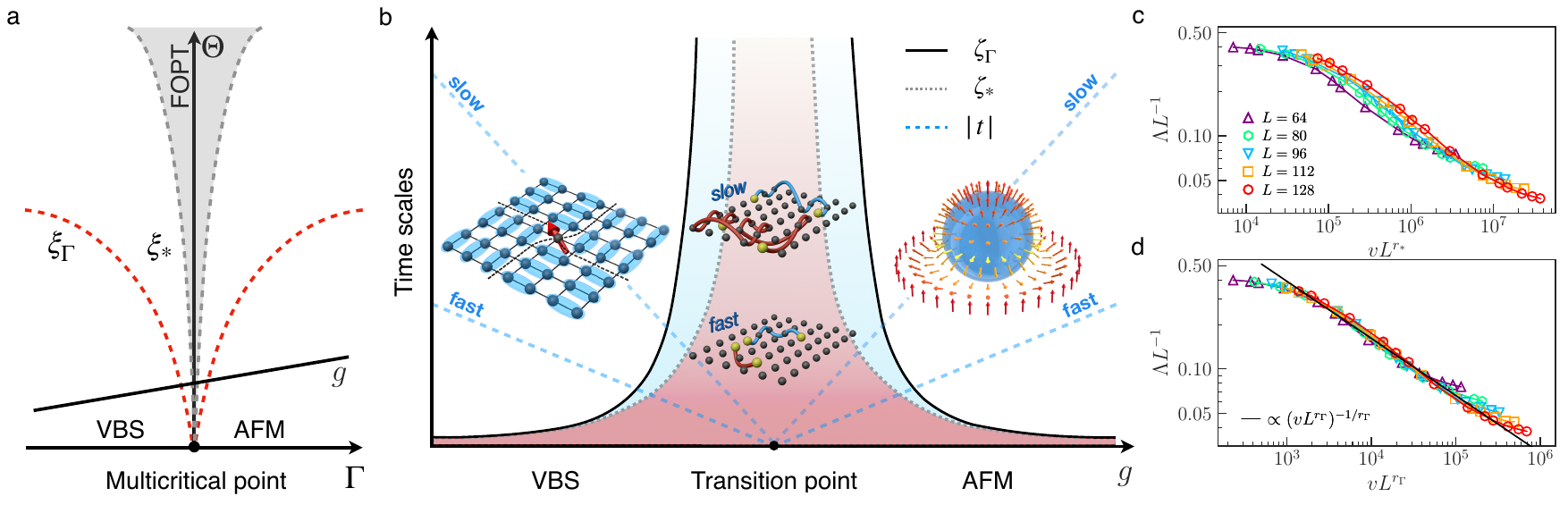}
  \caption{\textbf{Topological defects and divergent time scales.} {\bf a} Equilibrium phase diagram near the deconfined quantum multicritical point, with $\Theta$ and $\Gamma$ representing the two relevant fields. A first-order phase transition (FOPT) between the valence-bond-solid (VBS) and antiferromagnetic (AFM) phase occurs when crossing the vertical axis at $\Theta>0$, illustrated by the slanted line and corresponding to
    tuning a single parameter $g$ in a spin model. The red dashed curves separate the pseudocritical region with metastable bubbles of typical size $\xi_*$ from the critical region in which the fluctuations are dominated by deconfinement scale $\xi_\Gamma$. FOPT behavior with saturated length scales applies in the shaded region. {\bf b} Illustration of the dual asymmetric Kibble-Zurek scaling (DAKZS). As the tuning parameter $g$ approaches the critical point versus time, in addition to the pseudocritical correlation time $\zeta_*$, there is another time scale $\zeta_\Gamma$ characterizing the deconfinement process. The relevant time scale in DAKZS depends on the order parameter studied and the driving direction.
    The blue dashed lines indicate the temporal distance to the critical point for different driving rates.
    The topological defects in the two ordered phases are illustrated; vortices with $S=1/2$ cores (spinons, the red arrow in the left) and hedgehogs (monopoles, the blue sphere in the right) in the VBS and AFM phases, respectively. In the middle, deconfinement of two spinons (strings in a sampled quantum Monte Carlo configuration of an $S=1$ state) initially located at nearest-neighbor sites is illustrated for short and long annealing times. The dynamic scaling of the average string length $\Lambda$ violates the conventional Kibble-Zurek scaling with $\zeta_{v*}$ ({\bf c}), while satisfying the DAKZS with $\zeta_{v\Gamma}$ being the relevant time scale ({\bf d}). The line has a slope of $-1/r_{\Gamma}$, as expected for the scaling function when both $vL^{r_\Gamma}$ and $L$ are large. Log-log scale is used in {\bf c} and {\bf d}. Statistical errors here and in other figures are smaller than the symbol size.}
  \label{fig:conf}
\end{figure*}

Moreover, when relaxing the constraint of only one relevant operator with the symmetries of the Hamiltonian, numerical CFT bootstrap
studies \cite{Li2022jhep,Chester2024prl} also lend support to the scenario of deconfined multicritical point with emergent SO$(5)$ symmetry,
with an SO$(5)$-singlet operator with $\Theta$ as its strength being relevant in addition to the traceless symmetric tensor
of strength $\Gamma$. The latter relevant field changes sign at the AFM-VBS transition when tuned by a single parameter in a microscopic
model, as illustrated in Fig.~\ref{fig:conf}a. Remarkably, the critical exponents of the multicritical point determined from bounds by
the conformal bootstrap method \cite{Chester2024prl} are in good agreement with those estimated from the quantum Monte Carlo (QMC) simulations
of the $JQ$ model, which was also expanded with a second parameter to tune the $\Theta$ field \cite{Takahashi2024}. Rich scaling properties,
governed by two relevant directions combined with the dangerously irrelevant perturbations present in the lattice models, were revealed beyond
the expectations from conventional LGW theory~\cite{Shao2016,Takahashi2024}.

Recent studies with the fuzzy-sphere regulation of the SO$(5)$ nonlinear sigma model with a topological
Wess-Zumino-Witten term \cite{Zhou2024prx,Chen2024prb} also yield results consistent with the above discussions, though they have been interpreted
as supporting the non-unitary CFT scenario \cite{Zhou2024prx}.

Universal critical phenomena are manifested not only in equilibrium states but also in nonequilibrium processes. We here study dynamical
aspects of DQC through quantum annealing of the $2$D $JQ$ model in imaginary time, solving the corresponding Schr\"odinger equation using
a QMC approach and uncovering novel scaling behaviors stemming from slow equilibration of topological defects at the AFM--VBS transition.
We will here assume a multicritical real CFT description, but our conclusions only rely on near-criticality and should hold even for a complex CFT.

Within the conventional LGW description, when tuning a parameter $g = \pm v|t|$ (a reduced temperature or some parameter of the Hamiltonian)
versus time $t$, such that $g=0$ is a classical or quantum critical point, the velocity ($v$) dependence of the correlation length and
other critical properties are then described by the Kibble-Zurek scaling (KZS) mechanism
\cite{Kibble1976jpa,Zurek1985nat,Polkovnikov2011rmp,Dziarmaga2010}, which we outline first.

During the initial stage of an annealing process, the temporal distance $|t|$ to the critical point should exceed the relaxation
time $\zeta \propto \xi^z$ ($z$ being the dynamic critical exponent and $\xi$ being the correlation length). The system remains in equilibrium with correlation length
$\xi(t)\sim |g(t)|^{-\nu}$ until entering the impulse stage when $|t|<\zeta(t)$, falling out of equilibrium with the correlation
length frozen at~\cite{Kibble1976jpa,Zurek1985nat,Polkovnikov2011rmp,Dziarmaga2010}
\begin{equation}
\xi_v\sim v^{-1/r},~~~r\equiv z+1/\nu.
\label{xivdef}
\end{equation}
As a consequence, physical quantities are controlled by $\xi_v$ instead of $\xi$ from that point until $g=0$.

More precisely, because of Eq.~(\ref{xivdef}), the order parameter $P$ for a system of size $L$ satisfies an extended finite-size scaling
form that can be written as \cite{Zhong2005prb,DeGrandi2011prb,Liu2014prb,Huang2014prb}
\begin{equation}
\label{m1}
P^2(v,L)=L^{-(d-1+\eta)}f_P(vL^{r},gL^{1/\nu}),
\end{equation}
where $\eta$ is the anomalous dimension. The scaling function $f_P$ depends on the direction of annealing;
driving the system from an initially ordered or disordered state. In a sufficiently large system ($L \gg \xi_v$) annealed to $g=0$, the scaling
function must develop power-law behavior in $vL^{r}$ such that $L$ is either eliminated (when starting in the ordered phase) or
$P^2 \propto L^{-d}$ (approaching from the disordered phase in $d$ dimensions) \cite{Liu2014prb,Huang2014prb},
\begin{equation}
\label{m2}
P^2(v,L) \sim \left \lbrace \begin{array}{ll} v^{(d-1+\eta)/r}, & ~g_0 > 0, \\ L^{-d}v^{(\eta-1)/r}, & ~g_0 < 0, \end{array} \right.
\end{equation}
if $g_0>0$ corresponds to an initially ordered state.

Applications of KZS range from cosmology to condensed matter and qubit arrays
\cite{Kibble1976jpa,Zurek1985nat,Lin2014natphy,Clark2016sci,Rysti2021prl,Rams2019prl,Keesling2019nat,King2023nat,Suzuki2024prl}.
While classical KZS has been studied extensively using simulated (Monte Carlo) annealing \cite{Gong2010njp,Liu2014prb,Huang2014prb,Liu2015pre},
testing predictions of quantum KZS \cite{Zurek2005prl,Dziarmaga2005prl,Polkovnikov2005prb} in models beyond one-dimension ($1$D) is in general challenging because solutions of the
time-dependent Schr\"odinger equation are limited to small system sizes \cite{Schmitt2022sciadv}. However, for models accessible to QMC simulations,
KZS with identical exponents can be studied by annealing in imaginary time \cite{DeGrandi2011prb} as outlined in Appendix~\ref{app:methods} and illustrated for quantum Ising models in Appendix~\ref{app:tfim}. The reason is that, for both real-time and imaginary-time annealing from an initial ground state,
KZS governs the joint contribution from both the ground state and the low-energy excited states near the critical point. As the only tuning parameter describing the extent of departure from the equilibrium state, the driving velocity $v$ provides a natural characteristic quantity to describe the annealing dynamics in both real- and imaginary-time direction. In addition, the scaling dimension of $v$ is the same for both real- and imaginary-time annealing.
Accordingly, the KZS in imaginary time shares the same scaling form and exponents with that in real time, though details of the the scaling function
differ.

Here we demonstrate that conventional KZS does not apply to the annealing dynamics across the FOPT near the multicritical point of DQC. Using the $2$D $JQ$ model with AFM and VBS ground states as an example, as illustrated in Fig.~\ref{fig:conf}b, imaginary-time annealing shows that the dynamic scaling properties of the order parameters are affected by both the deconfinement time scale $\zeta_\Gamma\propto \xi_\Gamma^z$ with $\xi_\Gamma$ being the deconfinement length scale and a pseudocritical time scale $\zeta_*\propto \xi_*^z$ with $\xi_*$ being an emergent length scale associated with the bubble size of metastable state near the FOPT. The dominant time scale for the order parameter depends on the direction of the driving, i.e., from which ordered state the annealing process is started. The directionality
further obeys a duality stemming from the topological defects in the two phases. We construct an extension of KZS that we will refer to as dual asymmetric KZS (DAKZS). Here, it should be noted that the metastability aspect probed by the DAKZS is completely different from the KZS crossover investigated recently at classical first-order transitions \cite{Suzuki2024prl}.

Below we first define the spin model and give a brief review on the equilibrium critical properties, then outline our DAKZS ansatz and demonstrate its validity using imaginary-time annealing; see Appendix~\ref{app:methods} for technical details. As a contrast, we also discuss classical simulated annealing of a three-dimensional ($3$D) clock model to confirm that the symmetry crossover length in the ordered phase does not affect the dynamic scaling of the order parameter. Therefore, conventional KZS describes the annealing dynamics in this case.

~\\
{\bf RESULTS}\\
{\bf $JQ$ model and equilibrium length scales}\\
We study the annealing dynamics of the spin-$1/2$ $JQ_3$
model~\cite{Sandvik2007prl,Lou2009prb}, with the Hamiltonian
\begin{equation}
\label{eq:hamiltonian}
H=-J\sum_{\langle ij\rangle}P_{ij}-Q \hskip-2mm \sum_{\langle ijklmn\rangle} \hskip-2mm  {P_{ij}P_{kl}P_{mn}},
\end{equation}
where $P_{ij}\equiv 1/4-{\bf S}_{i}\cdot{\bf S}_{j}$ are singlet projectors and $\langle ij\rangle$ and $\langle ijklmn\rangle$ denote, respectively,
nearest-neighbor sites and three nearest-neighbor pairs in both horizontal and vertical columns on the $2$D square lattice. We take $Q=1$ as the unit of energy and
change $J$ versus imaginary time $t$ under different driving protocols, starting from the ground state of the system at some initial value $J_0$ stochastically projected out of a trial state. The choice of initial state is discussed further in Appendix~\ref{app:velo}.

Like the $JQ_2$ model \cite{Sandvik2007prl} (with two singlet projectors instead of three in the $Q$ term), it was shown that the $JQ_3$ model hosts
an AFM ground state for large $J$ with the order parameter ${\bf M}$ defined as
\begin{equation}
{\bf M}\equiv \frac{1}{L^2} \sum_{r_x,r_y}(-1)^{r_x+r_y}{\bf S}_r,
\end{equation}
in which $r_x$ and $r_y$ are the coordinates of lattice sites in $x$ and $y$ directions, respectively, and a columnar VBS order $(D_x,D_y)$ for small $J$ with the order parameter defined as
\begin{equation}
D_a\equiv \frac{1}{L^2}\sum_{r_x,r_y}(-1)^{r_a}({\bf S}_r\cdot{\bf S}_{r+\hat{a}}), ~~~\hat{a} = \hat x, \hat y,
\end{equation}
in which $\hat x$, $\hat y$ are the unit lattice vectors along $x$ and $y$ directions, respectively. The transition point is at $J_c\approx 0.671$~\cite{Lou2009prb,Shu2022prl} and here $g=J-J_c$. Note that a more
robust VBS order for $J=0$ exists in the $JQ_3$ model than that in the $JQ_2$ model.

It was recently confirmed that the AFM-VBS transition in the both $JQ_3$ and $JQ_2$ models is a weakly FOPT
~\cite{Chen2024prl,DEmidio2023sp,Takahashi2024,Deng2024prl,DEmidio2024prl}.
In addition, the FOPT in these models was shown to be close to a deconfined multicritical point exhibiting emergent SO$(5)$ symmetry~\cite{Chester2024prl,Takahashi2024,Deng2024prl,DEmidio2024prl}. This scenario provides a synergistic resolution to the long-standing enigmas of the
perceived DQC violations in microscopic models~\cite{Jiang2008jst,Chen2013prl,Harada2013prb,Ma2018prb,Wang2017prx}, and additional support has been gained from the remarkable consistency between CFT bootstrap studies \cite{Li2022jhep,Chester2024prl} and the most recent QMC simulations \cite{Takahashi2024}.

In contrast to a usual critical point, there exist multiple characteristic length scales at the AFM-VBS FOPT near the multicritical
DQC point, which we summarize here first in order to properly account for the time scales entering out of equilibrium:

\begin{enumerate}[1)]
\item Length scales associated with relevant fields. (i) $\xi_\Theta\propto |\Theta|^{-\nu_\Theta}$ with the critical exponent $\nu_\Theta=1.376(5)$~\cite{Chester2024prl,Takahashi2024} related to the strength of the SO$(5)$-singlet operator $\Theta$~\cite{Chester2024prl}. (ii) $\xi_\Gamma\propto |\Gamma|^{-\nu_\Gamma}$ with the critical exponent $\nu_\Gamma=0.632(4)$~\cite{Chester2024prl,Takahashi2024} related to the strength of traceless symmetric tensor operator $\Gamma$. Since $\nu_\Gamma$ works as a primary correlation length exponent in the CP$^1$ model~\cite{Chester2024prl}, whose building blocks are spinons and gauge fields, $\xi_\Gamma$ describes the typical length scale of the fractionalized degrees of freedom, namely the correlation length of gauge field excitation in AFM phase or the deconfinement length in VBS phase~\cite{Senthil2004,Levin2004prb}. We will refer to $\xi_\Gamma$ as the deconfinement scale.

\item Length scales induced by dangerously irrelevant lattice perturbations. For instance, in square lattice, where the VBS breaks the Z$_4$ lattice symmetry, the critical dimension for the leading dangerously irrelevant lattice perturbation, called $\Omega$, is $y_4=-0.723(11)$~\cite{Chester2024prl,Takahashi2024}. This results in a length scale $\xi_\Gamma'\propto |\Gamma|^{-\nu_\Gamma'}$ with $\nu_\Gamma'=\nu_\Gamma(1+|y_4|/p)$ describing the crossover length scale from emergent U$(1)$ to Z$_4$ lattice symmetry in the VBS phase along the $\Gamma$-direction~\cite{Takahashi2024}. $\xi_\Gamma'$ is manifested in the angular fluctuations of the order parameter, which can be used to quantify the evolution of the Z$_4$ symmetry deep inside the VBS to emergent U$(1)$ symmetry close to the transition \cite{Shao2020prl,Takahashi2024}. However, here the value of $p$ is still under debate \cite{Patil2021prb,Takahashi2024}.

\item A pseudocritical length scale $\xi_*\propto |\Gamma|^{-\nu_*}$~\cite{Takahashi2024}. Unlike the length scales in 1) and 2), which are solely governed by the multicritical point, $\xi_*$ is interpreted as the typical size of metastable bubbles of the second phase inside one of the ordered phases~\cite{Takahashi2024}, reflecting an unusual aspect of the near-DQC FOPT. In proximity of the multicritical point, $\xi_*$ exhibits scaling properties controlled by the DQC, with $\nu_*$ obeying the scaling law $\nu_*=\nu_{\Theta}/(1/\nu_{\Gamma}+|y_4|+1)=0.416(2)$. The value is close to that previously estimated for
what was assumed to be the conventional correlation length exponent, $\nu \approx 0.45$ \cite{Nahum2015prx,Zhao2020cpl}.
\end{enumerate}

Although for very large $L$ and small $|g|$, characteristic FOPT behaviors with finite length scales must ultimately apply [in the shaded region of Fig.~\ref{fig:conf}a], universal scaling behaviors controlled by length scales up to hundred or more lattice spacings can be observed in the $JQ_2$ and $JQ_3$ models
\cite{Takahashi2024}. Some additional remarks are as follows:

\begin{enumerate}[1)]
\item In the $JQ$ model, the distance to the transition point $g$ can generally be decomposed into $\Gamma$ and $\Theta$ components. However, since $\nu_\Theta$ is much larger than $\nu_\Gamma$, the $\Gamma$-direction typically dominates and $|g|\simeq \Gamma$, except when the parameter is tuned along the FOPT line~\cite{Takahashi2024}.
\item $\xi_*$ is governed by $\Gamma$, $\Theta$, and another dangerously irrelevant perturbation, called $\Omega_1$, which corresponds to the subleading lattice perturbation with the scaling dimension of $(y_4-1)$~\cite{Takahashi2024}. Different from $\Omega$ which only works in the VBS phase, $\Omega_1$ influences both AFM and VBS orders equally by generating an energy barrier between the AFM and VBS domains. This intrinsically prevents the AFM-VBS transition from being a conventional spin-flip FOPT in SO$(5)$ superspin space~\cite{Takahashi2024}, as no metastable domain can form in the latter. Thus, $\Omega_1$ plays an essential role in the emergence of $\xi_*$.
\item Based on the fact that $\nu_\Gamma$ is larger than $\nu_*$, an intuitive bubble-in-bag physical picture was proposed~\cite{Takahashi2024}. Near the multicritical point, $\xi_\Gamma$, which describes the typical distance between the topological defects---spinons in the VBS phase and space-time hedgehog singularities in the AFM phase---spans a bag in one of the phases, called phase A. Since the topological defects in one phase correspond to the order parameter in the other phase, these topological defects can serve as the seeds (nucleation cores) for the other phase, called phase B. Accordingly, the fluctuating bubble (metastable domain) of phase B, with size $\xi_*$, can form around the topological defects in the bag more easily than it would directly from the purely ordered region of phase A~\cite{Takahashi2024}.
\item For small $L$ (higher energy scale) or larger $|g|$ (far from the FOPT), the metastable domain cannot form, and both AFM and VBS correlations are described by $\xi_\Gamma$ with critical exponent $\nu_\Gamma$. In contrast, for larger $L$ and smaller $|g|$, the metastable domain can form, and both AFM and VBS correlations are instead described by $\xi_*$ with critical exponent $\nu_*$. This crossover explains the puzzle of the drift of the correlation length exponents.
\end{enumerate}

~\\
{\bf Multiple KZS length scales}\\
The conventional KZS length scale Eq.~(\ref{xivdef}) governs the out-of-equilibrium annealing dynamics on approach to a conventional critical point~\cite{Kibble1976jpa,Zurek1985nat,Polkovnikov2011rmp,Dziarmaga2010}. In contrast, for the annealing dynamics across the FOPT near the multicritical point, one may expect several dynamic length scales, associated with the equilibrium scales
  discussed above and illustrated in Fig.~\ref{fig:conf}b. These are
\begin{enumerate}[1)]
\item A dynamic deconfinement length scale
\begin{equation}
\xi_{v\Gamma}\sim v^{-1/r_\Gamma},~~~r_\Gamma\equiv z+1/\nu_\Gamma,
\label{xivpdef}
\end{equation}
which is obtained by comparing the time distance $|t|$ and the deconfinement time scale $\zeta_\Gamma\sim \xi_\Gamma^z$ in analogy with the procedure to obtain Eq.~(\ref{xivdef}). Accordingly, $\xi_{v\Gamma}$ should characterize the deconfinement scale of the DQC under driving.

\item A dynamic pseudocritical length scale
\begin{equation}
\xi_{v*}\sim v^{-1/r_*},~~~r_*\equiv z+1/\nu_*,
\label{xivp1def}
\end{equation}
corresponding to the length scale at $|t|\simeq \zeta_*\sim \xi_*^z$.

\item A dynamic length scale induced by dangerously irrelevant lattice symmetry
\begin{equation}
\xi_{v\Gamma}'\sim v^{-1/r_\Gamma'},~~~r_\Gamma'\equiv z+1/\nu_\Gamma',
\label{xivp2def}
\end{equation}
corresponding to the length scale at $|t|\simeq \zeta_\Gamma'\sim \xi_\Gamma'^z$. The value of $\nu_\Gamma'$ has still not been determined~\cite{Takahashi2024}.

\end{enumerate}

We here study annealing dynamics under changing $J$, which implies that the dominant scales is governed by the field $\Gamma$ when
crossing the transition at a very small value of the SO(5) singlet field $\Theta$, as in Fig.~\ref{fig:conf}a.

The appearance of multiple dynamic length scales qualitatively reshapes the KZS mechanism, prompting the formulation of a generalized KZS as follows: For a general observable $Y$, the finite-size scaling in the vicinity of the transition at $g=0$ can be expressed as
\begin{equation}
\label{eq:operator}
Y(v,g,L)=L^{-\Delta}f_Y(v L^{\tilde{r}},g L^{1/\tilde{\nu}}),
\end{equation}
where $\Delta$ is the scaling dimension of $Y$ and $\tilde{r}$ corresponds to $r_\Gamma$, $r_*$ or $r_\Gamma'$, depending on $Y$ and the annealing direction (in a way that will be determined below); similarly, $\tilde{\nu}$ should be $\nu_\Gamma$, $\nu_*$ or $\nu_\Gamma'$. With $\tilde{r}=r$ and $\tilde{\nu}=\nu$, this generalized KZS ansatz reduces to the conventional KZS form Eq.~(\ref{m1}).

In the following, we will focus on the dynamic scaling behaviors of the square of the order parameters after highlighting the dynamic deconfinement process. For these quantities, we will find that $\xi_{v\Gamma}'$ with its unknown exponent $\nu_\Gamma'$ does not enter. Therefore, we mainly discuss the effects induced by $\xi_{v\Gamma}$ and $\xi_{v*}$.

We will reach the limit where $L$ exceeds both $\xi_{v\Gamma}$ and $\xi_{v_*}$, so that the system is essentially in the thermodynamic limit at $g=0$ and power
laws developing for large scaling argument $v L^{\tilde{r}}$ in Eq.~(\ref{eq:operator}) lead to forms analogous to Eq.~(\ref{m2}). We will be far from
the limit $v L^{\tilde{r}} \to 0$
for large $L$, thereby circumventing the ultimate non-critical aspects of the FOPT \cite{Nahum2015prx,Shao2016}.

~\\
{\bf Dynamic deconfinement}\\
The central notion in the DQC is the fractionalization of quasi-particles at the critical point. Before presenting the general DAKZS ansatz, we first confirm the dynamic deconfinement scale $\xi_{v\Gamma}$. To this end, we investigate the deconfinement process of an
initially localized (at two neighboring sites) triplet excitation embedded in the VBS background, starting the annealing process of the $JQ_3$ model deep
inside its VBS phase at $J=0$ and ending at $J=J_c$.
We monitor the size $\Lambda$ of the spinon pair defined via the strings connecting unpaired spins in a singlet
background in the valence-bond basis in the $S=1$ sector \cite{Shao2016,Tang2013prl}. As shown in Fig.~\ref{fig:conf}c, $\Lambda L^{-1}$ data for different
annealing velocities and system sizes graphed versus $vL^{r_*}$ do not collapse well onto a common scaling function $f_\Lambda(vL^{r_*})$. In contrast, as shown in
Fig.~\ref{fig:conf}d, good data collapse is achieved by assuming $\Lambda = Lf_\Lambda (vL^{r_\Gamma})$. Then, for large $vL^{r_\Gamma}$ (and sufficiently
large $L$), $f_\Lambda\propto (vL^{r_\Gamma})^{-1/r_\Gamma}$, $\Lambda\propto \xi_{v\Gamma}\propto v^{-1/r_\Gamma}$, and we can also write
\begin{equation}
\label{eq:scalingfla}
\Lambda(v,L)=v^{-1/r_\Gamma} \tilde f_\Lambda(vL^{r_\Gamma}),
\end{equation}
confirming that the deconfinement process is controlled by $\xi_{v\Gamma}$ rather than $\xi_{v*}$.

~\\
{\bf DAKZS of the order parameters}\\
The equilibrium numerical results show that a single pseudocritical length scale $\xi_*$ with the critical exponent $\nu_*$ governing the divergence (until FOPT saturation)
of both the AFM and VBS correlation lengths in both phases near $J_c$ for large system size~\cite{Takahashi2024,Zhao2020cpl}. Therefore, one might expect the KZS of the AFM and VBS order parameters controlled by $\xi_{v*}$ to apply for both $D^2(v,L)$ and $M^2(v,L)$, irrespective of the phase from which the critical point is approached.

In contrast, here we propose the DAKZS ansatz, where both the pseudocritical dynamic length scale $\xi_{v*}$ and the deconfinement dynamic length scale $\xi_{v\Gamma}$ enter the scaling of the order parameters. Moreover, the scaling also depends on the direction in which the system is driven through the transition point according to a duality principle connecting the topological defects in the two phases.

\begin{figure*}[t]
\centering
  \includegraphics[width=\linewidth,clip]{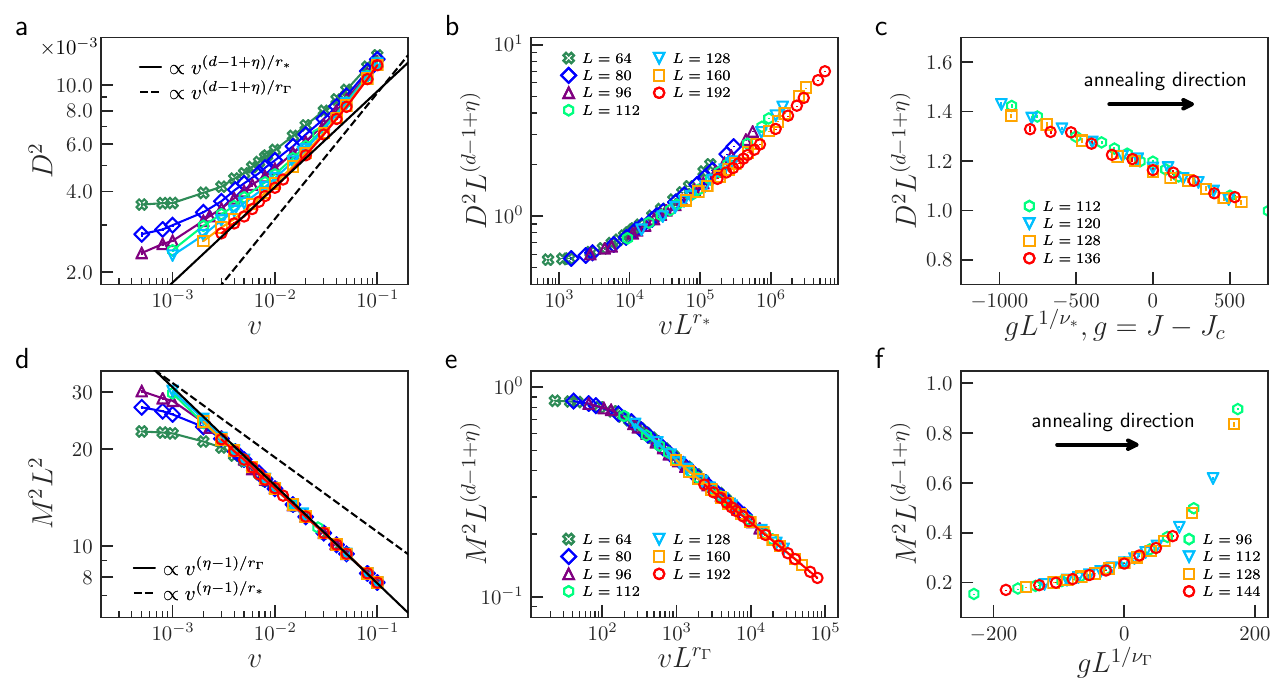}
  \caption{\textbf{Annealing from the valence-bond-solid phase.} {\bf a} Dependence of the valence-bond-solid order parameter on the driving velocity $v$ at the critical point for different system sizes $L$ [legends in {\bf b}]. The
    solid line shows the power law $D^2\propto v^{(d-1+\eta)/r_*}$ fitted to the $L=192$ data, while the slope of the dashed line corresponds to
    the exponent when $r_* \to r_{\Gamma}$. {\bf b} Data collapse with $v$ rescaled by $L^{r_*}$ and $D^2$ by $L^{(d-1+\eta)}$. {\bf c} For fixed $vL^{r_*}=5\times 10^{4}$, the scaled
    $D^2$ data collapse for large $L$ when $g$ is rescaled by $L^{1/\nu_*}$. {\bf d} The size-scaled antiferromagnetic order parameter vs $v$. The solid line shows
    $M^2L^2\propto v^{(\eta-1)/r_\Gamma}$ and the dashed line has slope corresponding to $r_\Gamma\to r_*$. {\bf e} Data collapse with $v$
    rescaled by $L^{r_\Gamma}$ and $M^2$ by $L^{(d-1+\eta)}$. {\bf f} Data collapse at fixed $vL^{r_\Gamma}=5\times 10^{3}$ with $g$ rescaled by $L^{1/\nu_\Gamma}$. Both order parameters share the same anomalous dimension $\eta=0.214(8)$~\cite{Takahashi2024}. Linear scale is used in {\bf c} and {\bf f} while log-log scale is used in other panels.
  }
  \label{fig:vbs}
\end{figure*}

Figure \ref{fig:vbs} shows results for the squared order parameters at $J=J_{\rm c}$ after annealing from the VBS phase at $J=0$. The VBS
order parameter in Fig.~\ref{fig:vbs}a indeed exhibits scaling behavior consistent with $D^2\propto v^{(d-1+\eta)/r_*}$, as in the first line of Eq.~(\ref{m2}) but with $r \to r_*$, for small $v$ and sufficiently large $L$. 
Figures~\ref{fig:vbs}b and \ref{fig:vbs}c show that the variables in the
scaling function are $vL^{r_*}$ and $gL^{1/\nu_*}$; thus $D^2$ satisfies the KZS form
\begin{equation}
\label{D1}
D^2(v,L)=v^{(d-1+\eta)/r_*}f_{D1}(vL^{r_*},gL^{1/\nu_*}).
\end{equation}

In contrast, Fig.~\ref{fig:vbs}d shows that the square of the AFM order parameter in the VBS phase is governed by the deconfinement length scale $\xi_{v\Gamma}$, having the
form $M^2L^2\propto v^{(\eta-1)/r_{\Gamma}}$ (indicated by the solid line), as on the second line of Eq.~(\ref{m2}) but with $r \to r_\Gamma$. The velocity
scaling is highly consistent across all the system sizes before the crossover toward the $v$-independent equilibrium value. Conventional KZS governed by
the exponent $r_*$ (slope indicated by the dashed line) can be excluded. As shown in Figs.~\ref{fig:vbs}e and \ref{fig:vbs}f, $r_\Gamma$ is also
established as the correct exponent in more comprehensive data analysis at both $g=0$ and $g \not= 0$. Thus $M^2$ satisfies the KZS form
\begin{equation}
\label{M1}
M^2(v,L)=L^{-d}v^{(\eta-1)/r_\Gamma}f_{M1}(vL^{r_\Gamma},gL^{1/\nu_\Gamma}).
\end{equation}

To explain the underlying mechanism behind different dominant length scales for AFM and VBS order parameters, we combine the KZS principle with the bubble-in-bag scenario. Under annealing from the VBS side, the KZS asserts that the topological defects, which are the spinons here, can emerge with a typical distance $\xi_{v\Gamma}$. These defects, acting as nucleation seeds, allow AFM metastable domains to form with a typical size $\xi_{v*}$. Accordingly, the reduction of the VBS order parameter is controlled by $\xi_{v*}$. Moreover, the topological defects are not independent, of each other but rather are entangled on the scale $\xi_{v\Gamma}$. Thus, adjacent spinons, residing on the cores of vortices and antivortices of VBS domains, should exhibit opposite SO$(3)$ vectors~\cite{Levin2004prb}. As a consequence, the adjacent AFM bubbles are also not independent with random orientations, but are correlated with each other on the scale of $\xi_{v\Gamma}$. Therefore, $M^2$, which integrates the spin correlations over the entire system size, is characterized by the scale $\xi_{v\Gamma}$. This dynamic asymmetry between the order parameters is one of the defining aspects of DAKZS.

\begin{figure*}[t]
\centering
  \includegraphics[width=\linewidth,clip]{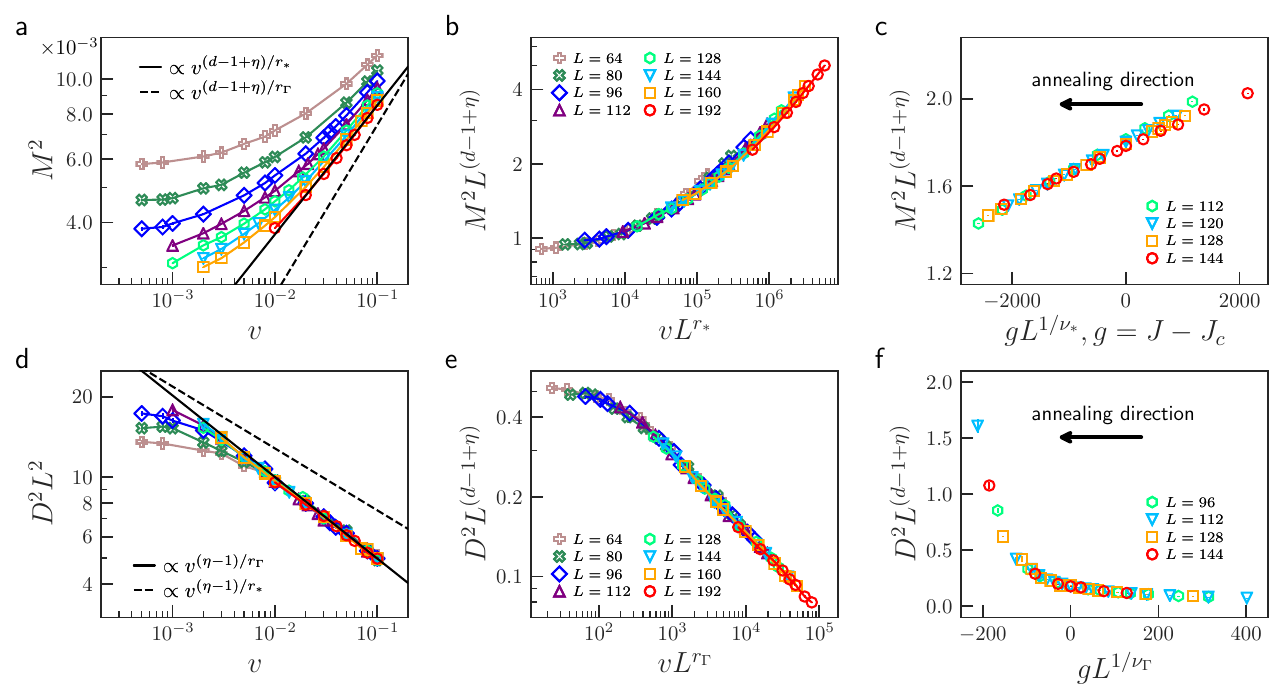}
  \caption{\textbf{Annealing from the antiferromagnetic phase.} {\bf a} The antiferromagnetic order parameter $M^2$ vs $v$ at the critical point for different system sizes $L$
    [legends in {\bf b}]. The solid line shows the form $M^2\propto v^{(d-1+\eta)/r_*}$ with amplitude matching the $L=192$ data, while the dashed line shows
    the slope when $r_* \to r_\Gamma$. {\bf b} Data collapse  when $v$ is rescaled by $L^{r_{*}}$ and $M^2$ by $L^{(d-1+\eta)}$. {\bf c} Data at fixed $vL^{r_*}=2\times 10^{5}$
    collapse for large $L$ when $g$ is rescaled by $L^{1/\nu_{*}}$. {\bf d} The size-scaled valence-bond-solid order parameter, with the form $D^2L^2\propto v^{(\eta-1)/r_\Gamma}$
    shown by the solid line; the dashed line corresponds to the exponent when $r_\Gamma \to r_*$. {\bf e} Data collapse when $v$ is
    rescaled by $L^{r_\Gamma}$ and $D^2$ by $L^{(d-1+\eta)}$. {\bf f} Scaled $D^2$ data vs $gL^{1/\nu_\Gamma}$ for fixed $vL^{r_\Gamma}=5\times 10^{3}$.
    Linear scale is used in {\bf c} and {\bf f} while log-log scale is used in other panels.
  }
  \label{fig:afmr}
\end{figure*}

A second remarkable aspect of DAKZS is the duality manifested when the annealing direction is reversed. To observe this duality, we consider an annealing process starting in the equilibrium AFM state at $J=1.2$, then gradually reducing $J$ at different velocities until $J_{\rm c}$ is reached. Figure~\ref{fig:afmr} shows results for the two order parameters analyzed in a manner analogous to Fig.~\ref{fig:vbs}. Here the AFM
order parameter obeys the KZS form with the exponent $r_*$,
\begin{equation}
\label{M2}
M^2(v,L)=v^{(d-1+\eta)/r_*}f_{M2}(vL^{r_*},gL^{1/\nu_*}),
\end{equation}
while the VBS order parameter satisfies the KZS form controlled by $r_\Gamma$,
\begin{equation}
\label{D2}
D^2(v,L)=L^{-d}v^{(\eta-1)/r_\Gamma}f_{D2}(vL^{r_\Gamma},gL^{1/\nu_\Gamma}).
\end{equation}

The physical interpretation here is similar to the previous case with a VBS ordered initial state. The hedgehog monopoles generated by annealing in the AFM background have a typical size $\xi_{v\Gamma}$.
These monopoles are coherent due to quantum interference between their Berry phases~\cite{Senthil2004} and serve as seeds for the correlated metastable VBS domains with size $\xi_{v*}$, which determines the reduction of the AFM order. Moreover, the correlation between different VBS domains results in the square of VBS order parameter being governed by $\xi_{v\Gamma}$.

Note that, in equilibrium~\cite{Takahashi2024} a balance should be reached between the growth of both AFM and VBS orders from correlated bubbles and the reduction of them from opposite bubbles. In this situation, $\xi_*$ dominates the correlation of both orders since it is shorter than $\xi_\Gamma$. In contrast, in the annealing dynamics, the order corresponding to the initial state works as a background, with which the reduction dominates the main order parameter, while the increase dominates the opposite order parameter. As a result, the roles played by $\xi_{v*}$ and $\xi_{v\Gamma}$ differ from each order, giving rise to the asymmetric aspect of DAKZS~\cite{Takahashi2024}.

Owing to the emergent SO$(5)$ symmetry of the DQC point \cite{Nahum2015prl,Takahashi2020prr}, a duality between the AFM and VBS order parameters is expected in equilibrium. Our results reveal a dynamic duality between the topological defects, where the longer deconfinement scale controls not only the equilibration of spinons in the VBS phase but also the hedgehog defects in the AFM phase, extending the emergent symmetry to the nonequilibrium realm.

~\\
{\bf Clock model}\\
Here we illustrate that the scale $\xi_{v\Gamma}'$ related to the crossover from U$(1)$ to discrete Z$_4$ symmetry in the VBS phase does not affect the DAKZS of the square of order parameters. However, since the value of $\nu_{\Gamma}'$ is still under debate \cite{Patil2021prb}, we turn to the well-studied $3$D classical $q$-state clock models. For $q \ge 4$ they exhibit U$(1)$ $\to$ Z$_q$ crossover of the order parameter symmetry at a length scale $\xi'_q$ controlled by an exponent $\nu_q'=\nu(1+|y_q|/2) > \nu$ \cite{Oshikawa2000prb,Shao2020prl}, similar to the  U$(1)$ to Z$_4$ crossover in the VBS phase.

\begin{figure*}[t]
\centering
  \includegraphics[width=\linewidth,clip]{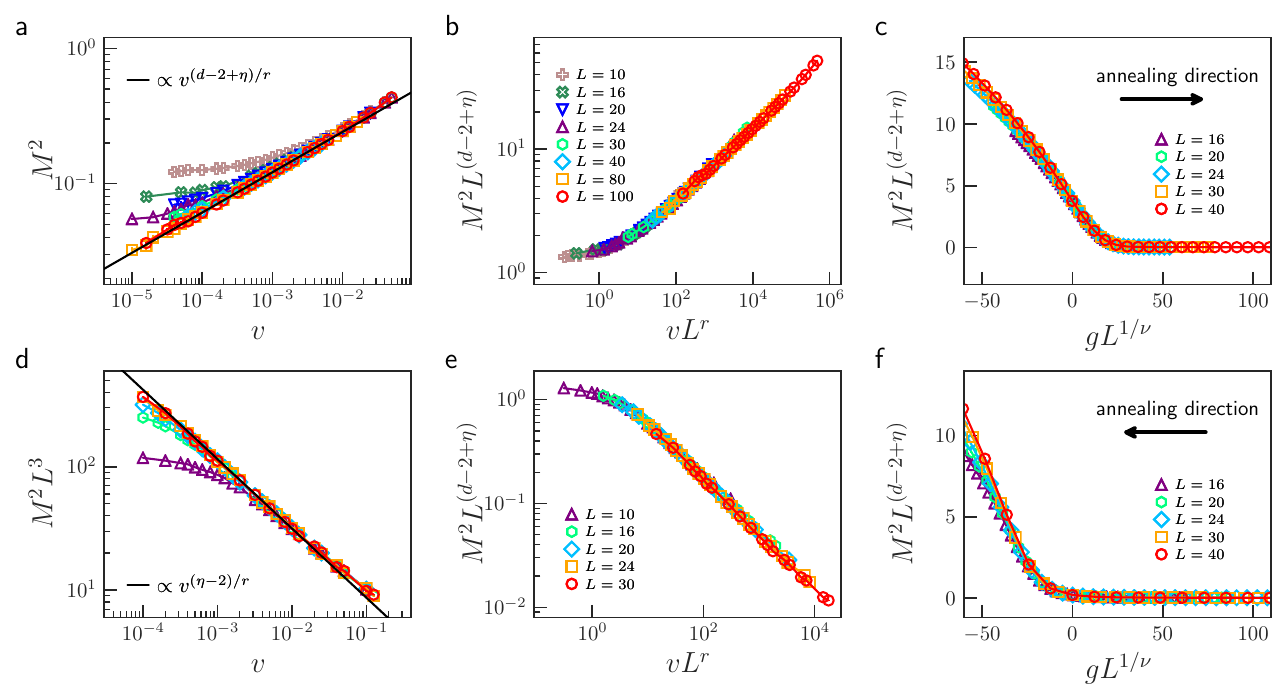}
  \caption{\textbf{Conventional Kibble-Zurek scaling in the classical $q=6$ clock model}.
    Upper row: Results for the average squared magnetization $M^2$ after linear simulated annealing to $T_c=2.20201$ \cite{Shao2020prl}
    from the ordered phase at $T_0=T_c-2$ (in energy unit). {\bf a} Dependence of $M^2$ on $v$ for different system sizes [legends in {\bf b}]. The
    line indicates the expected power law $\propto v^{(d-2+\eta)/r}$ with $3$D XY exponents $\eta=0.0380(4), z=2.0246(10)$~\cite{Chester2020jhep,Adzhemyan2022pa}. {\bf b} Collapse of the data of {\bf a} after rescaling as
    $M^2L^{(d-2+\eta)}$ vs $vL^{r}$. {\bf c} Data collapse in the vicinity of $g=0$ at fixed $vL^{r}=100$ and $g$ rescaled by $L^{1/\nu}$, where $g=T-T_0$.
    Lower row: Results at $T_c$ after annealing from the paramagnetic phase at $T_0=T_c+2$. {\bf d} Volume-scaled $M^2$ vs $v$ for different
    system sizes [legends in {\bf e}]. The line indicates the expected power law $\propto v^{(\eta-2)/r}$ for small $v$ and large $L$. {\bf e}
    Data collapse after rescaling to $M^2L^{(d-2+\eta)}$ vs $vL^{r}$. {\bf f} Data collapse with $g$ rescaled to $gL^{1/\nu}$ at fixed $vL^{r}=100$.
    Linear scale is used in {\bf c} and {\bf f} while log-log scale is used in other panels.
  }
  \label{fig:m2zq}
\end{figure*}

We study the $q=6$ clock model using classical simulated annealing. 
As shown in Fig.~\ref{fig:m2zq}, the squared order parameter is indeed fully
described by the KZS of Eq.~(\ref{m1}) with $3$D XY exponents, with no sign of any second scale controlled by $\nu'_6=1.52 > \nu=0.6717(1)$ \cite{Shao2020prl,Campostrini2001prb,Campostrini2006prb,Chester2020jhep}.
As discussed in detail in Appendix~\ref{app:clock}, the scale governed by $\nu'_6$ does appear in a quantity directly probing the Z$_q$ symmetry, which can be analyzed
with a KZS-inspired generalization of the equilibrium two-length scaling formalism developed in Ref.~\cite{Shao2020prl}.

~\\
{\bf DISCUSSION}\\
Our numerical simulations
have uncovered additional richness of the DQC phenomenon that was not anticipated in previous works. The dynamic
scaling ansatz that we call DAKZS can be summarized by replacing $r=z+1/\nu$ on the second line of Eq.~(\ref{m2}) by $r_\Gamma=z+1/\nu_\Gamma$ when $P$ is the order
parameter that is not long-ranged in the phase from which the annealing is started, becoming long-ranged on the other side of the transition. Calling said
(dis)order parameter $P_{\rm dis}$, its scaling form including also the behavior slightly inside the phases can be written as a specific case
of the general form Eq.~(\ref{eq:operator}) combined with analogue of the second of the asymptotic forms in Eq.~(\ref{m2});
\begin{equation}
P_{\rm dis}^2(v,L)=L^{-d}v^{(\eta-1)/r_\Gamma}\tilde f_{P}(vL^{r_\Gamma},gL^{1/\nu_\Gamma}).
\label{pdis}
\end{equation}
Equilibrium critical scaling $P_{\rm dis}^2\propto L^{-(d-1+\eta)}$ can be recovered from this form when $g=0$ and $v\rightarrow 0$ (by the scaling function
$\tilde f_{\rm P}$ developing a power-law form).

For much larger system sizes than those studied here, a FOPT with nonuniversal behaviors [in the shaded region around the FOPT line in Fig.~\ref{fig:conf}a] takes place in equilibrium. However, for annealing dynamics at finite velocity, the universal dynamic scaling behaviors described here are still expected as long as the dynamic length scales are smaller than those of the FOPT (as in the standard KZM \cite{Suzuki2024prl}). This assertion is further confirmed by the good scaling Figs.~\ref{fig:vbs}a, d and \ref{fig:afmr}a, d in the regions of larger $v$, where the finite-size effects should indeed be weaker on
account of the smaller dynamic length scales.

The dynamics of topological defects expose the deconfinement dynamic length scale in the case of DQC, which correspond to the $S=1/2$ cores of the VBS vortices and antivortices in the VBS phase and the space-time hedgehogs in the AFM phase~\cite{Levin2004prb}, while the formation of the fluctuating metastable bubbles around the topological defects has the pseudocritical dynamic length scale $\xi_{r_*}$. It is the interplay between these two scales that gives rise to the rich scaling behaviors described by the DAKZS ansatz~\cite{Senthil2006prb,Wang2017prx}.

Dualities and topological defects are at the heart of many exotic quantum phases and transitions of interest in studies of exotic quantum
matter \cite{Wang2017prx,Wang21,Lu21,Christos2023pnas}. Our DAKZS ansatz opens a window beyond KZS to emergent topological degrees of freedom through their
dynamics out of equilibrium. Noting that arrays of controllable Rydberg atoms provide a platform to realize both exotic quantum phases and
driven dynamics \cite{Keesling2019nat}, our scaling approach can potentially be directly exploited in that context in the near future.

~\\
{\bf Acknowledgements}\\
We would like to thank T. Senthil for valuable discussions.
This work was supported by the National Natural Science Foundation of China (Nos. 12222515, 12075324 to S.Y. and 12104109 to Y.R.S.), Key Discipline of Materials Science and Engineering, Bureau of Education of Guangzhou (No. 202255464 to Y.R.S.), a startup fund at Tulane University (to S.K.J.) and the Simons Foundation (No. 511064 to A.W.S.).
The authors would like to thank National Supercomputer Center in Guangzhou for providing high-performance computational resources and Guangzhou Ginpie Technology Co., Ltd. for collaboration in building a high-performance computing cluster for this project.
Some test cases were run on the Shared Computing Cluster managed by Boston University's Research Computing Services.

\appendix
\setcounter{equation}{0}
\renewcommand\theequation{A\arabic{equation}}
\setcounter{figure}{0}
\renewcommand\thefigure{A\arabic{figure}}

\section{Methods}
\label{app:methods}
\subsection{Quantum annealing}

The numerical quantum annealing results were obtained using the nonequilibrium quantum Monte Carlo (NEQMC) method \cite{DeGrandi2011prb},
which is a further development of the projector QMC method. Implementing Schr{\"o}dinger dynamics for a time dependent
Hamiltonian in imaginary time, it reproduces the same critical behavior (exponents) as in real time when annealing toward a critical point
\cite{DeGrandi2011prb,DeGrandi2013jpcm}. We here introduce the NEQMC method \cite{DeGrandi2011prb,Liu2013prb} and provide results for benchmark cases in Appendix~\ref{app:tfim}.

For a given imaginary-time (for which we now use the symbol $\tau$ instead of $t$ in the preceding sections)
dependent Hamiltonian $H(\tau)$, the evolution of a state obeys the analogue of
Schr{\"o}dinger dynamics, $|\psi(\tau)\rangle=U(\tau)|\psi(\tau_0)\rangle$ (up to an unimportant normalization constant), where
$|\psi(\tau_0)\rangle$ is the initial state and
\begin{equation}
U(\tau)=T_\tau\exp{\left[-\int_{\tau_0}^{\tau}{\rm d}\tau'H(\tau') \right]},
\end{equation}
is the Euclidean time evolution operator, with $T_\tau$ imposing time ordering.

In the NEQMC method, $U(\tau)$ is expanded in a power-series and applied to the initial state $|\psi(\tau_0)\rangle$, giving
\begin{equation}
  \begin{aligned}
    |\psi(\tau_{\rm a})\rangle=\sum_{n=0}^{\infty}&\int_{\tau_0}^{\tau_{\rm a}}{\rm d}\tau_{n} \int_{\tau_0}^{\tau_n}{\rm d}\tau_{n-1} \cdots \int_{\tau_0}^{\tau_3}{\rm d}\tau_{2} \int_{\tau_0}^{\tau_2}{\rm d}\tau_{1} \\
& \times  \left[-H(\tau_{n})\right]\cdots\left[-H(\tau_{1})\right]|\psi(\tau_0)\rangle,
\end{aligned}
\label{m:ints1}
\end{equation}
where we now denote the final annealing time $\tau_{\rm a}$ (which with $\tau_0=0$ is the total annealing time).
After inserting additional time integrals over unit operators $H_0$ at $m-n$ locations, with $m$ representing a truncation of the infinite series,
and splitting the Hamiltonian into bond (or other lattice units, e.g., the six-spin cells of the $Q_3$ interaction) operators,
\begin{equation}
  H=-\sum_{b=1}^{N_b}H_b,
\end{equation}
the state (whose normalization is irrelevant for Monte Carlo sampling) can be written as
\begin{equation}
  \begin{aligned}
    |\psi(\tau_{\rm a})\rangle=&\sum_{S_m}\frac{(m-n)!}{(\tau_{\rm a}-\tau_0)^{m-n}} \int_{\tau_0}^{\tau_{\rm a}}{\rm d}\tau_{n}
   \int_{\tau_0}^{\tau_n}{\rm d}\tau_{n-1} \cdots \\
    &~~~~ \cdots \int_{\tau_0}^{\tau_3}{\rm d}\tau_{2}\int_{\tau_0}^{\tau_2}{\rm d}\tau_{1}S_m |\psi(\tau_0)\rangle,
\end{aligned}
\label{m:ints2}
\end{equation}
in which $S_m$ denotes
the operator sequence,
\begin{equation}
  S_m=\prod_{i=1}^{m} H_{b_i}(\tau_{i}),
\end{equation}
and $n$ is the number of non-unit operators; $b_i\neq 0$. The factor before the integrals in Eq.~(\ref{m:ints2}) corrects for the integration
volume and number of possible insertions of the $m-n$ unit operators $H_0$. The truncation $m$ of the expansion must scale as
$m \propto L^d(\tau_{\rm a}-\tau_0)$ and is adapted self-consistently during the equilibration stage of the simulation, thus causing only
a vanishingly small truncation error (as in the finite-temperature stochastic series expansion method \cite{Sandvik2010review}).

To implement importance sampling of the normalization $Z=\langle\psi(\tau_{\rm a})|\psi(\tau_{\rm a})\rangle$,
the wave function is written in a basis $\{|\alpha\rangle\}$,
which for $S=1/2$ models with spin-isotropic interactions, like the $JQ$ models, can be conveniently taken to be the overcomplete valence-bond basis
\cite{Sandvik2007prl,Tang2011prl,Beach2006npb}. In this work we use the valence-bond basis for the $JQ_3$ model, while the transverse-field Ising models (see Appendix~\ref{app:tfim}) are simulated in the conventional spin-$z$ basis~\cite{DeGrandi2011prb,Liu2013prb}.

A full Monte Carlo sweep of the importance sampling procedure consists of local (diagonal) and global (loop) updates of $S_m$, $\tau_{m}$ and the basis
state $\{|\alpha\rangle\}$.
First, diagonal updates of $S_{m}$ and $\{|\alpha\rangle\}$ are carried out, with the time value $\tau_{m}$ fixed, similar to those in standard projector QMC and the stochastic series expansion QMC method~\cite{Sandvik2010review,Tang2011prl}.
In the next stage, loop updates are carried out, constructing loops that traverse the imaginary-time propagation, allowing global updates of the operator sequence and the states.
Since the probabilities of different types of operators now are associated with time values $\tau_{m}$ that enter in the probabilities for the diagonal operator updates in each propagation step in the imaginary-time direction, updates of the ordered time values are carried out successively, without changing $S_m$ and $\{|\alpha\rangle\}$.

To sample the time integrals in Eq.~(\ref{m:ints2}) efficiently, a multi-point update scheme is used. A random segment
$\{\tau_{i},...,\tau_{i+n_\tau}\}$ of time values is first chosen, with $n_{\tau}$ the length of the segment. Then random numbers in the range
$(\tau_{i-1},\tau_{i+n_{\tau}+1})$ are generated and ordered, leading to a new allowed and unbiased time set $\{\tau_{i}',...,\tau_{i+n_{\tau}}'\}$.
The Metropolis acceptance probability to replace the chosen time set with the new set can be easily obtained from Eq.~(\ref{m:ints2}). The length of
the segment $n_{\tau}$ is adjusted to maintain a reasonable mean acceptance rate, close to $1/2$.

Measurements of expectation values of operators $A$ are taken in the middle of the double-sided projection;
\begin{equation}
\langle A(\tau_{\rm a})\rangle=\frac{1}{Z}\langle\psi(\tau_{\rm a})|A|\psi(\tau_{\rm a})\rangle.
\end{equation}

In some systems (like those in Appendix~\ref{app:tfim}) it is convenient to use simple initial states that are trivially eigenstates
of appropriate starting Hamiltonians. With the $JQ$ models, there are no such simple eigenstates close to the AFM--VBS transition.
To generate initial states, we therefore carry out additional projections (as a part of the overall time evolution) with the Hamiltonian
fixed at its initial parameters before the time dependence is applied, i.e., in the initial stages of Eq.~(\ref{m:ints2}) up to some sufficiently
long time, the Hamiltonian is not dependent on $\tau$ (and the integrals can therefore also be eliminated). The final sampled state of these additional
projections then effectively serves as the initial state $|\psi(\tau_0)\rangle$ to which the successive driving (with the time dependent
Hamiltonian) is applied.

In the case of the $JQ_3$ model, we always fix $Q=1$, take the initial time in Eq.~(\ref{m:ints2}) as $\tau_0=0$ [when the time dependent interactions
is turned on after initial projection with fixed $H(J_0)$], and evolve with $\tau>0$. For the VBS initial state, $J_0=0$, the driving parameter
$J$ is increased following $J(\tau)=J_0+v\tau$, while with the AFM initial state, $J_0=1.2$, $J$ varies according to $J(\tau)=J_0-v\tau$. In calculations
with $g=0$, the final time value is $\tau_{\rm a}=|J_c-J_0|/v$, and when stopping at $g\neq 0$ the corresponding final $J$ value replaces $J_c$.
Note that $J(\tau)\geq 0$ must be satisfied in order to maintain the interactions antiferromagnetic---for negative $J$ the simulations are
affected by a sign problem.

The imaginary-time annealing should share the similar scaling form with the same critical exponents with the real-time annealing. To see this, one can expand the evolving wave-function in the basis of instantaneous eigenstates of the Hamiltonian and compare the coefficients of the excited states under annealing in real-time $t$ direction and imaginary-time $\tau$ direction~\cite{Polkovnikov2005prb,DeGrandi2011prb}. The initial state is always assumed to be the ground state for the initial parameter $g_0$ that is far from the critical point $g=0$.

For the real-time case, the coefficient for the $n$-th excited state reads~\cite{Polkovnikov2005prb}
\begin{equation}
  a_n(t)\simeq -\int_{g_0}^{g=0} dg' \langle n|\partial_{g'}|0\rangle \exp\left[-i\int_{g'}^{g=0}dg'' \frac{\Delta_{n0}(g'')}{v}\right],
\label{realt}
\end{equation}
in which $\Delta_{n0}(g)\equiv E_n(g)-E_0(g)$ is energy difference between the $n$-th eigenstate and the ground state of the instantaneous Hamiltonian $H(g)$.

For the imaginary-time case, the coefficient for the $n$-th excited state reads~\cite{DeGrandi2011prb,DeGrandi2013jpcm}
\begin{equation}
  \alpha_n(\tau)\simeq \int_{g_0}^{g=0} dg' \langle n|\partial_{g'}|0\rangle \exp\left[-\int_{g'}^{g=0}dg'' \frac{\Delta_{n0}(g'')}{v}\right].
\label{imaginarytau}
\end{equation}

By comparing Eqs.~(\ref{realt}) and (\ref{imaginarytau}), one finds that the difference is the constant coefficient of the argument in the exponential term, which is imaginary unit for real-time dynamics but unit for imaginary-time dynamics. However, universal scaling behaviors are contained in other shared variables, including: $g$, the transition matrix $\langle n|\partial_{g'}|0\rangle$, $\Delta_{n0}$, the energy difference $\Delta_{n0}(g)$, and the annealing velocity $v$. For the annealing dynamics, $v$ should be small. Therefore, when the initial state is the ground state, only low-energy excited states are involved. These states exhibit universal scaling behaviors controlled by the critical point. Moreover, in Eqs.~(\ref{realt}) and (\ref{imaginarytau}), $v$ is the only scaling variable characterizes how far the system is driven out of equilibrium, giving rise to the KZS. Accordingly, the scaling theory for both the real-time and imaginary-time dynamics should share a similar scaling form, with the same critical exponents, differing only in the specific details encoded within the scaling function.

\subsection{Classical simulated annealing}

The results of the clock model were obtained using simulated annealing Monte Carlo simulations. To implement classical simulated annealing, the simulations are performed in two stages; an initial equilibration stage followed by an annealing stage. 
During the initial equilibration stage, the simulations are carried out at a fixed initial temperature $T_0$ for a sufficiently long time, generating an equilibrium state to serve as the initial state. 
Next the evolution continues with the temperature varies with time according to $T=T_0\pm vt$ in each Monte Carlo sweep, where $v$ is the driving rate and $t$ is the simulation time. The total annealing time $t_a$ is determined by $t_a=|T_a-T_0|/v$, where $T_a$ is the target temperature. Measurements are taken at each desired temperature $T$, yielding results for different $T$ in a sing run.

\section{Conventional KZS in quantum Ising models}
\label{app:tfim}
\begin{figure*}[t]
\centering
  \includegraphics[width=\linewidth,clip]{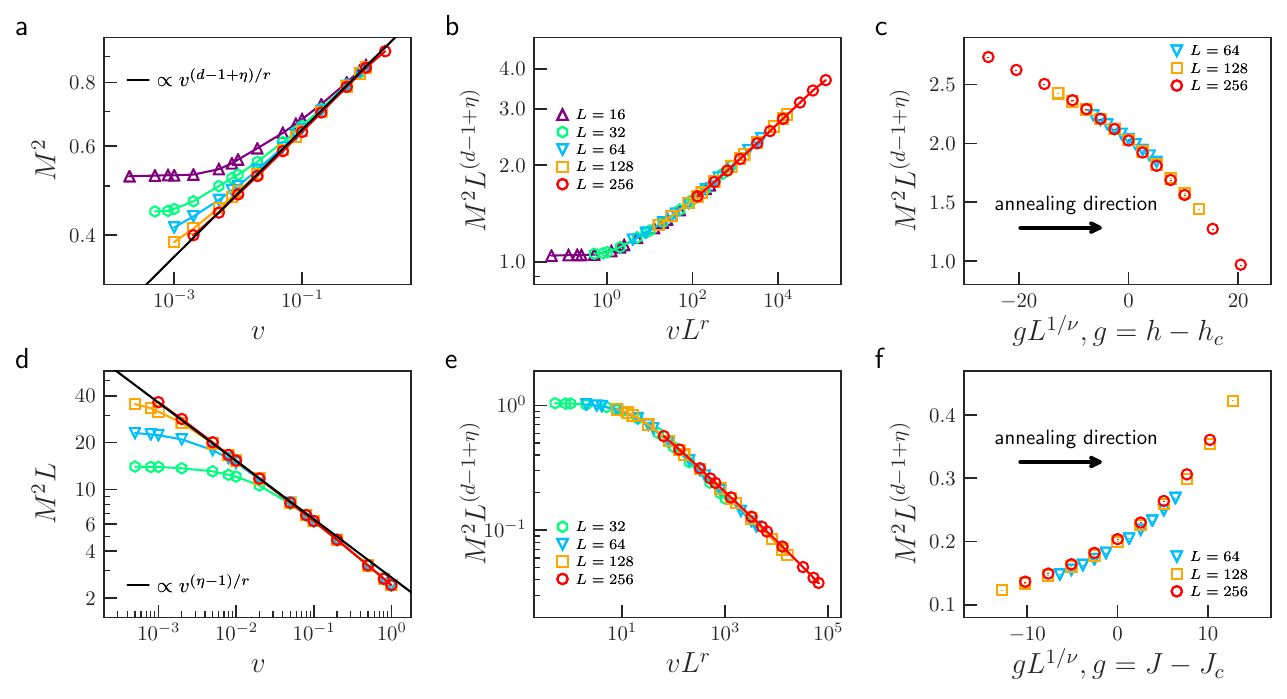}
  \vskip-3mm
  \caption{\textbf{Kibble-Zurek scaling in the one-dimensional quantum Ising model}.
    Upper row: Annealing from the ordered phase to the critical point. {\bf a} Dependence of $M^2$ on $v$ for $L=16$ to $256$
    [legends in {\bf b}]. The solid line indicates the power law $v^{(d-1+\eta)/r}$ applicable for large $L$ and small $v$.
    {\bf b} Rescaling $v$ and $M^2$ as $vL^r$ and $M^2L^{(d-1+\eta)}$,
    respectively, leads to data collapse as expected. {\bf c} Data collapse after rescaling $M^2$ versus $g$ in the driving process for $L=64$ to
    $256$ at fixed $vL^r=1000$. Lower row: driving from the disordered phase. {\bf d} Dependence of the size-scaled $M^2$ on $v$ at the critical point for $L=32$
    to $256$. The solid line shows the power law $v^{(\eta-1)/r}$ expected for large $L$ and small $v$.
    {\bf e} Rescaling $v$ and $M^2$ as $vL^r$ and $M^2L^{(d-1+\eta)}$, respectively,
    for data collapse. {\bf f} Rescaled $M^2$ versus $g$ for $L=64$ to $256$ at fixed $vL^r=1000$.
    Linear scale is used in {\bf c} and {\bf f}, while log-log scale is used in others.
    All results support the conventional Kibble-Zurek mechanism.
  }
  \label{fig:1dtfim}
\end{figure*}

\begin{figure*}[t]
\centering
  \includegraphics[width=\linewidth,clip]{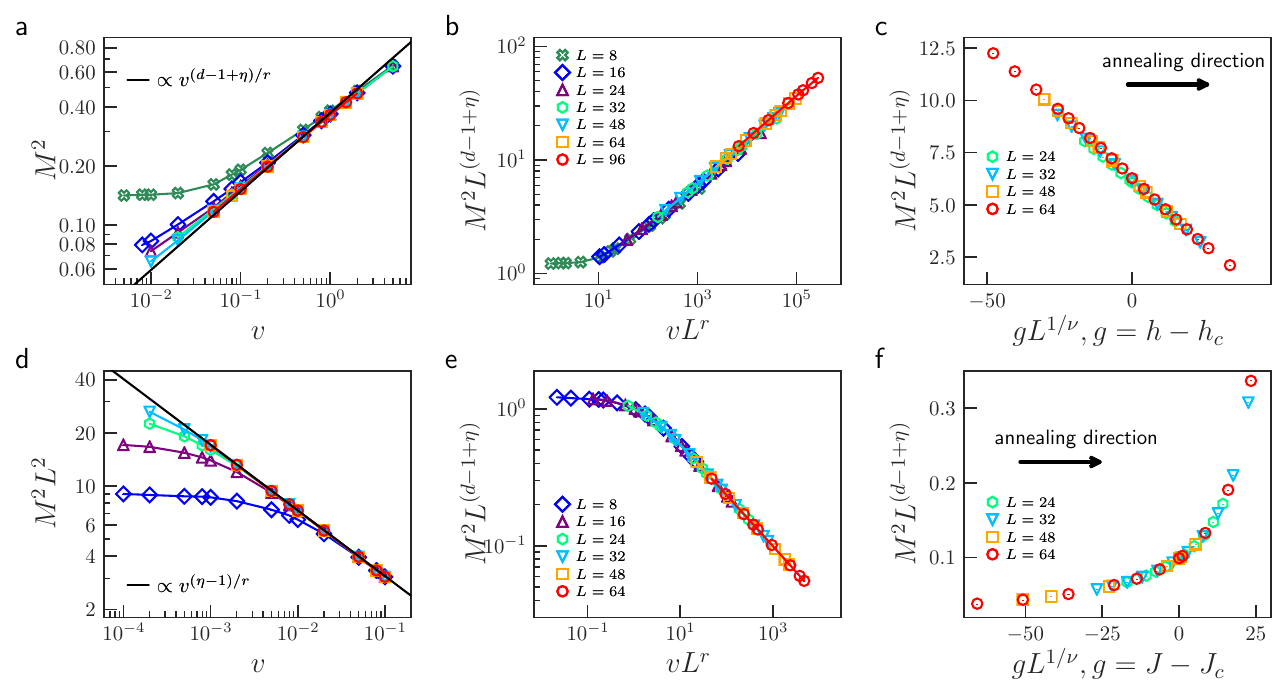}
  \vskip-3mm
  \caption{\textbf{Kibble-Zurek scaling in the two-dimensional quantum Ising model}. Upper row: driving from the ordered phase. {\bf a} Dependence of $M^2$ on
    $v$ at the critical point for $L=8$ to $96$. The solid line indicates the form $v^{(d-1+\eta)/r}$. {\bf b} Data collapse with $v$ and $M^2$ scaled
    by $vL^r$ and $M^2L^{(d-1+\eta)}$, respectively. The solid line shows the power law $(vL^{r})^x$ with the expected exponent $x=(d-1+\eta)/r$.
    {\bf c} Data collapse with the expected powers of $L$ in the neighborhood of the critical field at fixed $vL^{r}=1000$.
    Lower row: driving from the disordered phase. {\bf d} Dependence of the size-scaled $M^2$ on $v$ at the critical point for $L=8$ to $64$. The solid line shows
    the expected power law $v^{(\eta-1)/r}$ when $L$ is sufficiently large. {\bf e} Data collapse of $vL^r$ versus $M^2L^{(d-1+\eta)}$. {\bf f} Data
    collapse with the expected powers of $L$ in the neighborhood of the critical coupling at fixed $vL^{r}=1000$.
    Linear scale is used in {\bf c} and {\bf f}, while log-log scale is used in others.
    All results support the conventional Kibble-Zurek mechanism
    with the expected critical exponents.}
  \label{fig:2dtfim}
\end{figure*}

The quantum Ising model (the transverse-field Ising model)
serves as a testing ground for a variety of theories, experiments, and numerical simulations in the study of quantum critical
phenomena. The Hamiltonian defined on an arbitrary lattice with nearest-neighbor site pairs $\langle ij\rangle$ is
\begin{equation}
H=-J\sum_{\langle ij\rangle}\sigma^z_i\sigma^z_j-h\sum_{i}\sigma^x_i,
\end{equation}
where $\sigma^x_i$, $\sigma^z_i$ are Pauli matrices, $J>0$ is a ferromagnetic coupling, and $h$ is the strength of the transverse field.

The $1$D version of the model can be solved exactly~\cite{Suzuki1976}. It hosts a quantum phase transition between the ferromagnetic and paramagnetic
phases at the coupling ratio $(h/J)_c=1$. The critical exponents for the order parameter and the correlation length are $\beta=1/8$, $\nu=1$ and $\eta=1/4$,
respectively, following from the mapping to a $2$D classical model \cite{Sachdev1999,Sondhi1997rmp,Vojta2003rpp}.
The $2$D model has no rigorous analytical solution. Recent numerical simulations have placed the critical point at $(h/J)_c\approx3.04451$ on a square
lattice, and the universality class is again that of the classical model in one higher dimension, with the critical exponents $\beta\approx 0.3258$, $\nu\approx 0.6289$ and $\eta\approx 0.036$\cite{Shu2017prb,Liu2013prb,Hasenbusch2010prb}. The dynamic exponent is $z=1$ for both the $1$D and $2$D models,
reflecting the emergent Lorentz invariance of the model when mapped to the $D+1$ dimensional classical model.

The driven dynamics of the quantum Ising model obeys conventional KZS, and it has been confirmed in previous studies that imaginary-time
dynamics results in the same type of scaling behavior as in real time \cite{DeGrandi2011prb,DeGrandi2013jpcm}. Starting an annealing process
from the ordered side of the transition, the leading scaling form of the order parameter is
\cite{Gong2010njp,Zhong2011book,Liu2014prb,Huang2014prb}
\begin{equation}
\label{ms1}
M^2(v,L)=v^{(d-1+\eta)/r}f_{M1}(vL^{r},gL^{1/\nu}),
\end{equation}
in which $r=z+1/\nu$, $g$ is the distance to the critical point, and $v$ the driving velocity. In contrast, starting from the paramagnetic
side, the scaling form is
\begin{equation}
\label{ms2}
M^2(v,L)=L^{-d}v^{(\eta-1)/r}f_{M2}(vL^{r},gL^{1/\nu}),
\end{equation}
though the form of the scaling function $f_{M2}$ is not the same as in Eq.~(\ref{ms1}). Note, however, that although Eqs.~(\ref{ms1}) and (\ref{ms2})
have different leading terms when expressed in the annealing velocity, both of them can be converted into the form
\begin{equation}
M^2(v,L)=L^{-(d-1+\eta)}f_{M3}(vL^{r},gL^{1/\nu}),
\label{b:kzs}
\end{equation}
by redefining $f_{M3}$ by extracting appropriate power laws of the argument $vL^{r}$ from the scaling function [as was argued in reverse in
the main text, Eqs.~(2) and (3)]. Either of the forms can be used for analyzing data for different velocities and system sizes.

We perform nonequilibrium quantum Monte Carlo (NEQMC) simulations \cite{DeGrandi2011prb} (also see Methods) of both the $1$D and $2$D quantum Ising models.
For the ordered initial state, the driving protocol is $h=h_0+v\tau$ with $J=1$ and $h_0=0$, such that the initial state can be prepared
by setting all spins aligned ferromagnetically in the $z$ direction. Here, the distance to the critical point is defined as $g\equiv h-h_c$.
Therefore, driving the system to the critical point corresponds to linearly varying $h$ versus $\tau$
from $0$ to the critical value $h_c=1$ \cite{Sachdev1999,Sondhi1997rmp,Vojta2003rpp} and $3.04451$~\cite{Shu2017prb,Liu2013prb} for the $1$D and
$2$D case, respectively.

For the disordered initial state, instead of driving $h$, we choose $J=J_0+v\tau$ with $h=1$ and $J_0=0$, such that
the initial state can be easily prepared with the equal superposition of all spin-$z$ basis state;
\begin{equation}
|\psi(\tau_0)\rangle=\bigotimes_{i=1}^{L^d}(|\uparrow\rangle_i+|\downarrow\rangle_i).
\end{equation}
This initial state is sampled collectively along with the evolved states as dictated by the operators in the product $S_m$ in Eq.~(21) in Methods.

Driving the system to the critical point then means $J$ evolving from $J_0=0$ to $J_c=1$ and $0.32846$ for the $1$D and $2$D case, respectively.
Here the distance to the critical point is defined as $g\equiv J-J_c$. The previous NEQMC simulations only considered the paramagnetic initial
state \cite{DeGrandi2011prb,DeGrandi2013jpcm}. A ``two-way'' projection with mixed time boundaries was also considered with a slightly different
algorithm \cite{Liu2013prb}. The results presented here for both annealing directions extend to lower velocities and larger system sizes.

For the $1$D case, we show results starting from the ferromagnetic initial state in the upper row of Fig.~\ref{fig:1dtfim}. At the critical point
$g=0$, Fig.~\ref{fig:1dtfim}a shows that $M^2$ decreases as $v$ decreases. For a finite-size systems, as $v\rightarrow 0$, $M^2$ converges to its
equilibrium size-dependent critical value, $M^2 \propto L^{-(d-1+\eta)}$, as is explicit in Eq.~(\ref{b:kzs}) with an analytic scaling function $f_{M3}$.
For larger $v$, $M^2$ obeys $M^2\propto v^{(d-1+\eta)/r}$, as explicitly conveyed by Eq.~(\ref{ms1}), where the scaling function $f_{M2}$ must approach
a constant for large argument $vL^r$, as seen in Fig.~\ref{fig:1dtfim}a. The form with only $v$ dependence and
no size corrections from the scaling function requires $L$ to be sufficiently large, $L \gg \xi_v$. Furthermore, the correlation length should also
obey  $\xi_v \gg 1$, so that the system has evolved from the initial state and critical fluctuations have developed. After rescaling $v$ and $M^2$ as
$vL^r$ and $M^2 L^{(d-1+\eta)}$, respectively, we observe in Fig.~\ref{fig:1dtfim}b that all data collapse well, as predicted by Eq.~(\ref{b:kzs})
when $g=0$.

In Fig.~\ref{fig:1dtfim}c, we show that the rescaled curves of $M^2$ versus
$g$ in the vicinity of the critical point also collapse onto each other for a fixed value $vL^r$, in which case Eq.~(\ref{b:kzs}) again reduces to
a single-variable scaling form. These results demonstrate that, in the $1$D quantum Ising model, $M^2$ evolving from the ordered initial state
satisfies Eqs.~(\ref{ms1}) and (\ref{b:kzs}), as expected \cite{Zhong2011book,Gong2010njp,Huang2014prb,Liu2014prb}.

In the lower row of Fig.~\ref{fig:1dtfim}, we show results obtained when driving from the disordered paramagnetic phase. In Fig.~\ref{fig:1dtfim}d, at the critical point $M^2$ satisfies $M^2\propto v^{(\eta-1)/r}$ for small $v$ and large $L$, as predicted
by Eq.~(\ref{ms2}). After rescaling with the critical power of $L$, Fig.~\ref{fig:1dtfim}e shows that the data versus $vL^r$ for different system
sizes collapse onto a common scaling function to a high degree, again as expected from the KZS mechanism expressed in the form of Eq.~(\ref{b:kzs}).
In Fig.~\ref{fig:1dtfim}f, rescaled curves of $M^{2}$ versus $g$ for fixed $vL^r$ are shown to collapse almost perfectly
for $L=128$ and $L=256$, while for $L=64$ small scaling corrections are visible. These results confirm the scaling forms in Eqs.~(\ref{ms2})
and (\ref{b:kzs}) \cite{Gong2010njp,Zhong2011book,Huang2014prb}.

For the $2$D case, similar results are shown in Fig.~\ref{fig:2dtfim} and there is no need to comment further on these other than to conclude that
KZS works extremely well also in this case, with almost no scaling corrections visible.

\section{ Upper bound for the annealing velocity in the $JQ_3$ model}
\label{app:velo}

\begin{figure*}[!htbp]
\centering
  \includegraphics[width=\linewidth,clip]{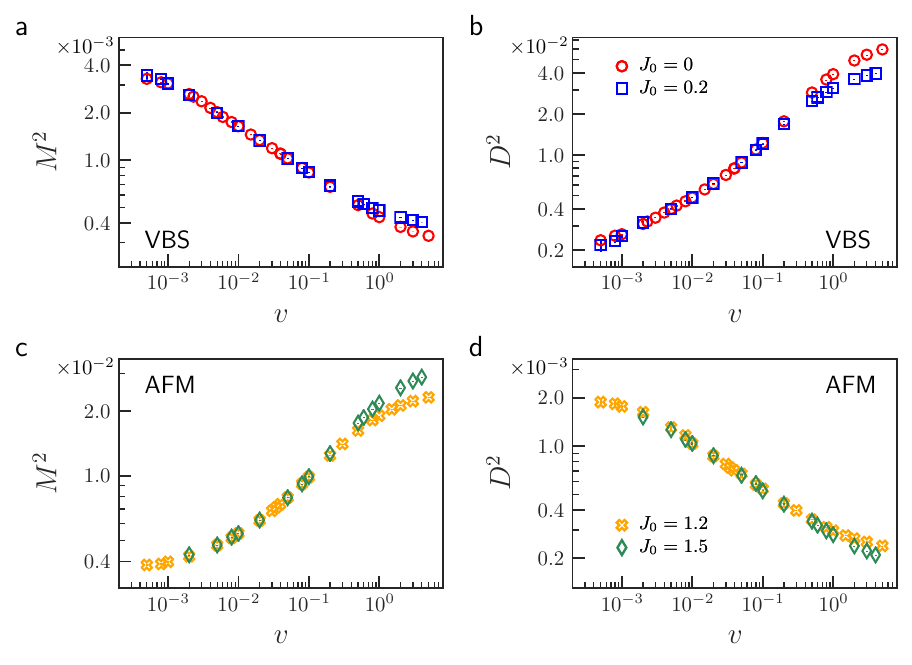}
  \vskip-3mm
  \caption{\textbf{Determination of upper bound of $v$}. Curves of $M^2$ in {\bf a} and $D^2$ in {\bf b} vs $v$ at the critical point starting from a
    valence-bond-solid (VBS) ground states with $J_0=0$ and $0.2$ for system size of $L=96$. For the two initial states, $M^2$ is statistically indistinguishable when
    $v<10^{-1}$, and the same applies to$D^2$ as well. Curves of $M^2$ in {\bf c} and $D^2$ in {\bf d} vs $v$ at the critical point starting from an antiferromagnetic (AFM)
    ground states with $J_0=1.2$ and $1.5$ for $L=96$. For $v<10^{-1}$, both $M^2$ and $D^2$ agree for the two different initial states. The results
    of $J_0=0$ and $1.2$ are from the main text.
  Log-log scale is used in all panels.}
  \label{fig:initstate}
\end{figure*}

KZS requires that the system evolves adiabatically before the critical region, ideally starting from an eigenstate of the initial Hamiltonian
(though this is strictly speaking not necessary in imaginary time, as long as the driving velocity is not too large given the initial parameters). Since a fast annealed
system is dominated by the initial conditions, there is some upper bound on $v$ above which scaling cannot be observed.

Here we determine the upper bound of the driving velocity, below which the effects induced by the initial state can be ignored when
the initial driven parameter is far from the critical point. For the VBS initial state, in the main text we use $J_0=0, Q=1$ and let $J=J_0+v\tau$.
Here we choose $J_0=0.2$ and perform NEQMC simulations on a system with $L=96$. As shown in Figs.~\ref{fig:initstate}a and \ref{fig:initstate}b,
for both $M^2$ and $D^2$, the differences between different initial states are small in the region of $v<10^{-1}$. Similarly, for the AFM initial
state, we compare the results starting from $J_0=1.5, Q=1$ with those starting from $J_0=1.2, Q=1$ (used in the main text) for $L=96$. As shown in Figs.~\ref{fig:initstate}c and \ref{fig:initstate}d, curves of $M^2$ and $D^2$ versus $v$ match each other for different initial states when
$v<10^{-1}$. Therefore, in the main text, the driving velocity $v$ is chosen to be smaller than $10^{-1}$ to guarantee that the effects induced
by the initial state are negligible.

In principle, it would always be better to start further away from the critical point. However, for a fixed velocity $v$, a larger range of
the annealing parameter implies slower simulations. While the projection of the initial state also takes time, it is faster than the annealing
stage. Our choice of initial condition represents a compromise between simulation efficiency and range of $v$ not affected by the initial state.

\section{Extended KZS in $3$D classical $Z_q$-clock models}
\label{app:clock}
\begin{figure*}[t]
\centering
  \includegraphics[width=\linewidth,clip]{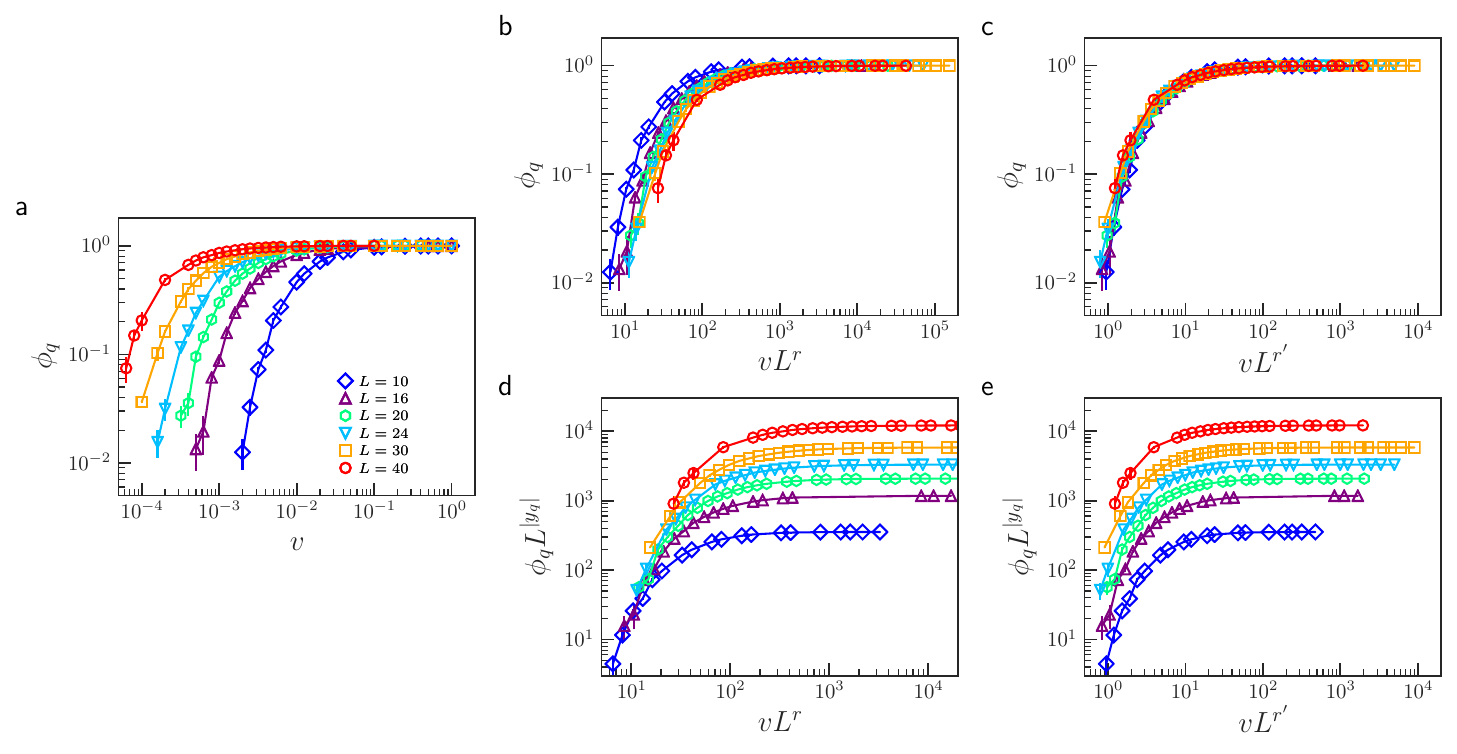}
  \vskip-3mm
  \caption{\textbf{Dynamic scaling of the angular order parameter $\phi_q$ ($q=6$); annealing from the ordered state.}
    {\bf a} Dependence of $\phi_q$ on $v$ for system of size $L=10$ to $40$, starting from the ordered state at $T_0=T_c-2=0.20201$.
    {\bf b} Rescaling $v$ as $vL^{r}$ does not collapse data for $\phi_q < 1$.
    {\bf c} Rescaling $v$ as $vL^{r'}$ leads to good data collapse when $\phi_q$ is close to $1$.
    {\bf d} Close to equilibrium, rescaling $\phi_q$ and $v$ as $\phi_q L^{|y_q|}$ and $vL^{r}$, respectively, leads to good data collapse for small values of $\phi_q$.
    {\bf e} No data collapse can be observed when rescaling with $vL^{r'}$.
    Log-log scale is used in all panels.
  }
  \label{fig:smphiq6fm}
\end{figure*}

Here we explore driven stochastic (Metropolis Monte Carlo) dynamics in the classical $Z_q$-clock model on the $3$D simple cubic
lattice. The Hamiltonian of this model is
\begin{equation}
H=-J\sum_{\langle i,j\rangle}{\rm cos}(\theta_i-\theta_j)-h\sum_{i}{\rm cos}(q\theta_i),
\label{zqhamil}
\end{equation}
in which $\theta\in [0,2\pi)$ and $\langle i,j\rangle$ denotes nearest neighbors. Although the system only has the discrete $Z_q$ symmetry, it was
shown that, for $q\geq 4$, the phase transition for $h \not= 0$ at temperature $T=T_c$ belongs to the $3$D $U(1)$ universality class, which means $h$ is
irrelevant at the critical point \cite{Jose1977prb,Pujari2015prb,Shao2020prl}. The critical exponents for the order parameter and the
correlation length are therefore $\beta=0.3486(1)$, $\nu=0.6717(1)$ and $\eta=0.0380(4)$ \cite{Campostrini2001prb,Campostrini2006prb,Chester2020jhep}. However,
in the ordered phase for $T<T_c$, $h$ is relevant, reducing the order parameter symmetry from $U(1)$ to $Z_q$. Accordingly, $h$ is categorized
as a dangerously irrelevant scaling variable
\cite{Oshikawa2000prb,Lou2007prl,Okubo2015prb,Campostrini2001prb,Leonard2015prl,Pujari2015prb,Hove2003pre,Hasenbusch2011prb,Shao2020prl,Patil2021prb,Chlebicki2022pre,Jose1977prb,Ueno1991prb,Chubukov1994prb,Miyashita1997jpsj,Zhitomirsky2014prb,Banerjee2018prl,Hasenbusch2019prb}. Thus, besides the usual correlation
length $\xi$, there is another relevant length scale $\xi'$ that characterizes the cross-over from U(1) to $Z_q$ symmetry of the order parameter (and
also the thickness of domain walls). This second length scale $\xi'$ is governed by the exponent $\nu' > \nu$ so that $\xi' \gg \xi$ upon approaching the
critical point.

In the following, we take $q=6$ as an example to explore the critical dynamics of the model. In this case, $\nu'=1.52$~\cite{Shao2020prl,Patil2021prb}.
We take the ``hard clock'' limit $h \to \infty$ in Eq.~(\ref{zqhamil}), implemented as a discrete set of $q=6$ sampled angles. The critical temperature in
this case in units of the coupling, $J=1$, is $T_c=2.20201$~\cite{Shao2020prl}. We perform Monte Carlo simulations using standard Metropolis
dynamics, in which case the dynamic exponent $z=2.0246(10)$~\cite{Adzhemyan2022pa}.
When starting the annealing process from the ordered state, the driving
protocol is given by $T=T_0+vt$ with $T_0=0.20201$, while from the disordered state we take $T=T_0-vt$ with $T_0=4.20201$ (i.e., the initial
temperatures represent $T_c/J \pm 2$). The annealing procedure in each case is repeated many (thousands or more) times and averages are taken
over the final configurations at $T_c$.

In analogy with the $JQ_3$ model in the main text, in the driven dynamics it is expected that there exists another characteristic velocity-limited
length scale $\xi'_v\propto v^{-1/r'}$ with $r'=z+1/\nu'$, in addition to the usual typical length scale $\xi_v\propto v^{-1/r}$ with $r=z+1/\nu$ in
KZS. As the results in Fig.~\ref{fig:m2zq} of the main text show, the magnitude of the order parameter (the magnetization) is not affected by the longer
scale and is described by conventional KZS. To reveal the second scale and develop an extended KZS formalism depending on it, we explore the annealing
dynamics of an angular order parameter $\phi_q$, defined by
\begin{equation}
\label{angop}
\phi_q\equiv\langle {\rm cos}(q\Theta)\rangle,
\end{equation}
with $\Theta={\rm arccos}(M_x/M)$, in which $M=(M_x^2+M_y^2)^{1/2}$ with
\begin{equation}
M_x=\frac{1}{L^3}\sum_i{\rm cos}(\theta_i),~~~ M_y=\frac{1}{L^3}\sum_i{\rm sin}(\theta_i).
\end{equation}
This angular order parameter and variations of it have been used in numerous equilibrium studies
\cite{Lou2007prl,Okubo2015prb,Pujari2015prb,Patil2021prb,Shao2020prl,Patil2021prb}. We here investigate the corresponding dynamical critical behavior.

Under external driving from the ordered initial state, Fig.~\ref{fig:smphiq6fm}a shows the dependence of $\phi_q$ on the driving velocity $v$ after
reaching the critical point. Figure \ref{fig:smphiq6fm}b shows that, after rescaling $v$ as $vL^{r}$ following KZS with the conventional
correlation-length exponent $\xi$, the rescaled curves do not match each other, whereas rescaling as $vL^{r'}$ in Fig.~\ref{fig:smphiq6fm}c leads to good
data collapse. In this case we have not applied any rescaling to $\phi_q$, as it should be regarded as a dimensionless quantity in the regime where it
is approaching the saturation value $\phi_q=1$ \cite{Shao2020prl} (as discussed in more detail further below). In contrast, when $\phi_q$ is small and
size-dependent, i.e., when it is governed by its critical scaling dimension $y_6= -2.55(6)$ \cite{Okubo2015prb,Shao2020prl}, it should be rescaled
correspondingly to observe data collapse.

As shown in Fig.~\ref{fig:smphiq6fm}d, with the critical rescaling we observe satisfactory data collapse for the smallest values of the velocity when
graphing versus $vL^r$ with the conventional KZS exponent $r=z+1/\nu$. Because of difficulties in obtaining good data for large systems at low velocity,
we cannot unambiguously claim that the asymptotic small-$vL^r$ scaling in Fig.~\ref{fig:smphiq6fm}d is better than that in Fig.~\ref{fig:smphiq6fm}c
for small $vL^{r'}$, but the data are at least consistent with $r$ controlling the near-equilibrium critical behavior. If we instead rescale as $vL^{r'}$,
there is no apparent data collapse at all, as seen in Fig.~\ref{fig:smphiq6fm}e. Thus, we conclude that the conventional KZS mechanism most likely applies
for small driving velocity, where $\phi_q$ decreases versus the system size, while for larger driving velocity, where $\phi_q$ is effectively dimensionless
because of the ordered and symmetry-broken initial state, an extended KZS mechanism applies that is governed by the exponent $r'=z+1/\nu'$.

The extended KZS mechanism at play in the clock model is still different from the DAKZS of the $JQ_3$ model, where there are two ordered states and a
duality between the two order parameters and their topological defects. We have not investigated any angular order parameter analogous to $\phi_q$
in the $JQ_3$ model, because of the excessive computational efforts required in order to collect sufficient data for proper analysis.

We next develop a scaling theory to understand the dynamic scaling of $\phi_q$ more formally. By generalizing the scaling form in Ref.~\cite{Shao2020prl} to
the non-equilibrium case, we find that the full scaling form of $\phi_q$ should be
\begin{equation}
\label{angopfull}
\phi_q(g,v,L)=L^{-|y_q|}f(gL^{1/\nu},gL^{1/\nu'};vL^r,vL^{r'}),
\end{equation}
in which $g\equiv T-T_c$. For a process stopping at the critical point, $g=0$, Eq.~(\ref{angopfull}) reduces to
\begin{equation}
\label{angopfull1}
\phi_q(g,v,L)=L^{-|y_q|}f(vL^r,vL^{r'}).
\end{equation}
Since the function $f$ should be analytic for small values of its arguments and $r>r'$, the argument $vL^r$ will dominate when $vL^{r'}\ll vL^r$. For
small velocities, Eq.~(\ref{angopfull1}) can therefore be approximated as
\begin{equation}
\label{angopfull2}
\phi_q(g,v,L)=L^{-|y_q|}j(vL^r),
\end{equation}
in which $j$ is another scaling function. Eq.~(\ref{angopfull2}) explains the scaling behavior found in Fig.~\ref{fig:smphiq6fm}d for the
smallest values of $v$ available for each system size.

For larger $v$, the scaling argument $vL^{r'}$ must also be included, and when $vL^r$ also becomes very large ($vL^{r} \gg vL^{r'}$) the function
$f$ should develop a power-law in it, resulting in the form
\begin{equation}
\label{angopfull3}
\phi_q(g,v,L)=L^{-|y_q|}(vL^r)^ak(vL^{r'}),
\end{equation}
with yet another scaling function $k$ and with the exponent $a$ to be determined. The above form should hold for any value of $vL^{r'}$ as
long as $vL^r$ is large and its power law has developed. When $\phi_q$ is close to $1$ (because of the initial condition), as demonstrated in Figs.~\ref{fig:smphiq6fm}b and \ref{fig:smphiq6fm}c, Eq.~(\ref{angopfull3}) should cross over to an explicitly dimensionless form;
\begin{equation}
\label{angopfull4}
\phi_q(g,v,L)=p(vL^{r'}).
\end{equation}
Eqs.~(\ref{angopfull3}) and (\ref{angopfull4}) now dictate that $k(vL^{r'})$ must satisfy
\begin{equation}
k(vL^{r'})=(vL^{r'})^b w(vL^{r'})
\label{kfunction}
\end{equation}
for some function $w(x)$ that approaches a constant for $x \to 0$. Combination of this scaling forms with Eq.~(\ref{angopfull3}) gives that
$b=-a$, in order to eliminate $v$ in front of the scaling function when $v \to 0$, and then to eliminate $L$ from Eq.~(\ref{angopfull3}) we must have
\begin{equation}
\label{angopfull6}
a=\frac{|y_q|}{r-r'}=\frac{|y_q|}{1/\nu-1/\nu'}.
\end{equation}
By using the values $\nu=0.6717$, $\nu'=1.52$, and $y_6=-2.55$ in Eq.~(\ref{angopfull6}), one finds that $a\approx 3.07$ for the $Z_6$ model.

\begin{figure}[!htbp]
\centering
  \includegraphics[width=0.8\linewidth,clip]{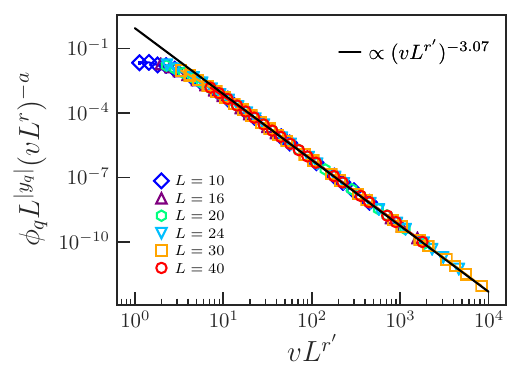}
  \vskip-3mm
  \caption{\textbf{Verification of the scaling theory for $\phi_q$.} After rescaling $v$ and $\phi_q$ as $vL^{r'}$ and $\phi_q L^{|y_q|} (vL^r)^{-a}$
    with the predicted exponent $a=3.07$ and with the other exponents set to their known values for the $q=6$ clock model, the  $\phi_q$ data for
    for different system sizes and velocities, confirming Eq.~(\ref{angopfull3}) with the derived value of $a$. Thus, the dynamics of $\phi_q$ is
    controlled by both $\xi_v$ and $\xi'_{v}$. The solid line indicates the power law $(vL^{r'})^{-a}$ with the predicted exponent $a=3.07$.
    Log-log scale is used.
  }
  \label{fig:smphiqa}
\end{figure}

We can test the above chain of arguments by using the predicted value of $a$ in Eq.~(\ref{angopfull3}). As shown in Fig.~\ref{fig:smphiqa}, rescaling
$v$ and $\phi_q$ as $vL^{r'}$ and $\phi_qL^{|y_q|} (vL^r)^{-a}$, respectively, the rescaled data collapse well for both the large-$v$ and small-$v$
regions. In addition, the slope of the common curve in Fig.~\ref{fig:smphiqa} is fully consistent with $-3.07$, indicating that the leading term in
$k(vL^{r'})$ in Eq.~(\ref{kfunction}) is $(vL^{r'})^b$ with $b=-a \approx -3.07$ as predicted above. In addition, when $vL^{r'}$ is small we observe
that the rescaled data begin to saturate, indicating that $k(vL^{r'})$ in Eq.~(\ref{angopfull3}) approaches a constant for small $vL^{r'}$. This behavior
is consistent with the function $k$ in Eq.~(\ref{angopfull2}) being analytical when the argument is taken to zero.

All these results confirm our derivations and demonstrate that both $\xi_v$ and $\xi'_{v}$ control the dynamics of $\phi_q$, i.e., that the driven dynamics
in the $Z_q$-symmetric clock model is beyond the conventional KZS mechanism. We point out that the data graphed as in Figs.~\ref{fig:smphiq6fm}c and d
represent limiting behaviors, while the scaling function $k$ obtained from the data collapse Fig.~\ref{fig:smphiqa} represent the complete behavior
together with Eq.~(\ref{angopfull3}) and the expression for the exponent $a$ in Eq.~(\ref{angopfull6}).

In Fig.~\ref{fig:m2zq} in the main text, we demonstrated that the order parameter $M^2$ is always controlled by $\xi_v$ for driven dynamics from both the ordered
phase and the disordered phase. There is a velocity regime in which $M^2\propto v^{(d-2+\eta)/r}$, as shown in Fig.~\ref{fig:m2zq}a, and the scaling collapse
in Figs.~\ref{fig:m2zq}b and \ref{fig:m2zq}c show that the dimensionless variables in the scaling function are $vL^{r}$ and $gL^{1/\nu}$. These results demonstrate
that the evolution of $M^2$ from the ordered initial phase satisfies Eq.~(\ref{ms1}). In addition, after annealing from a disordered initial state,
Fig.~\ref{fig:m2zq}d shows that $M^2$ satisfies $M^2L^3 \propto v^{(\eta-2)/r}$ until the cross-over to the equilibrium behavior. Again, Figs.~\ref{fig:m2zq}e
and \ref{fig:m2zq}f show that $v$ and $g$ should be rescaled as $vL^{r}$ and $gL^{1/\nu}$, respectively. Thus, the evolution of $M^2$ obeys conventional
KZS for both annealing directions and is not affected by the presence of the longer length scale.

The driven dynamics of the $Z_q$-clock model has been realized in experiments as reported in Ref.~\cite{Lin2014natphy}. There, the scaling of the defect
density was shown to obey the conventional KZS, as the behavior of the order parameter $M^2$ shown here. According to our present analyses, it would be
quite instructive to further explore the driven dynamics of the angular order parameter $\phi_q$ in this system.

\bibliographystyle{naturemag}
\bibliography{ref}

\begin{thebibliography}{10}
\expandafter\ifx\csname url\endcsname\relax
  \def\url#1{\texttt{#1}}\fi
\expandafter\ifx\csname urlprefix\endcsname\relax\def\urlprefix{URL }\fi
\providecommand{\bibinfo}[2]{#2}
\providecommand{\eprint}[2][]{\url{#2}}

\bibitem{Senthil2004}
\bibinfo{author}{Senthil, T.}, \bibinfo{author}{Vishwanath, A.},
  \bibinfo{author}{Balents, L.}, \bibinfo{author}{Sachdev, S.} \&
  \bibinfo{author}{Fisher, M. P.~A.}
\newblock \bibinfo{title}{{Deconfined Quantum Critical Points}}.
\newblock \emph{\bibinfo{journal}{Science}} \textbf{\bibinfo{volume}{303}},
  \bibinfo{pages}{1490--1494} (\bibinfo{year}{2004}).

\bibitem{Levin2004prb}
\bibinfo{author}{Levin, M.} \& \bibinfo{author}{Senthil, T.}
\newblock \bibinfo{title}{{Deconfined quantum criticality and N\'eel order via
  dimer disorder}}.
\newblock \emph{\bibinfo{journal}{Phys. Rev. B}} \textbf{\bibinfo{volume}{70}},
  \bibinfo{pages}{220403} (\bibinfo{year}{2004}).

\bibitem{Senthil2006prb}
\bibinfo{author}{Senthil, T.} \& \bibinfo{author}{Fisher, M. P.~A.}
\newblock \bibinfo{title}{{Competing orders, nonlinear sigma models, and
  topological terms in quantum magnets}}.
\newblock \emph{\bibinfo{journal}{Phys. Rev. B}} \textbf{\bibinfo{volume}{74}},
  \bibinfo{pages}{064405} (\bibinfo{year}{2006}).

\bibitem{Sandvik2007prl}
\bibinfo{author}{Sandvik, A.~W.}
\newblock \bibinfo{title}{{Evidence for Deconfined Quantum Criticality in a
  Two-Dimensional Heisenberg Model with Four-Spin Interactions}}.
\newblock \emph{\bibinfo{journal}{Phys. Rev. Lett.}}
  \textbf{\bibinfo{volume}{98}}, \bibinfo{pages}{227202}
  (\bibinfo{year}{2007}).

\bibitem{Melko2008prl}
\bibinfo{author}{Melko, R.~G.} \& \bibinfo{author}{Kaul, R.~K.}
\newblock \bibinfo{title}{{Scaling in the Fan of an Unconventional Quantum
  Critical Point}}.
\newblock \emph{\bibinfo{journal}{Phys. Rev. Lett.}}
  \textbf{\bibinfo{volume}{100}}, \bibinfo{pages}{017203}
  (\bibinfo{year}{2008}).

\bibitem{Lou2009prb}
\bibinfo{author}{Lou, J.}, \bibinfo{author}{Sandvik, A.~W.} \&
  \bibinfo{author}{Kawashima, N.}
\newblock \bibinfo{title}{{Antiferromagnetic to valence-bond-solid transitions
  in two-dimensional $\text{SU}(N)$ Heisenberg models with multispin
  interactions}}.
\newblock \emph{\bibinfo{journal}{Phys. Rev. B}} \textbf{\bibinfo{volume}{80}},
  \bibinfo{pages}{180414} (\bibinfo{year}{2009}).

\bibitem{Chen2013prl}
\bibinfo{author}{Chen, K.} \emph{et~al.}
\newblock \bibinfo{title}{{Deconfined Criticality Flow in the Heisenberg Model
  with Ring-Exchange Interactions}}.
\newblock \emph{\bibinfo{journal}{Phys. Rev. Lett.}}
  \textbf{\bibinfo{volume}{110}}, \bibinfo{pages}{185701}
  (\bibinfo{year}{2013}).

\bibitem{Harada2013prb}
\bibinfo{author}{Harada, K.} \emph{et~al.}
\newblock \bibinfo{title}{{Possibility of deconfined criticality in SU($N$)
  Heisenberg models at small $N$}}.
\newblock \emph{\bibinfo{journal}{Phys. Rev. B}} \textbf{\bibinfo{volume}{88}},
  \bibinfo{pages}{220408} (\bibinfo{year}{2013}).

\bibitem{Nahum2015prx}
\bibinfo{author}{Nahum, A.}, \bibinfo{author}{Chalker, J.~T.},
  \bibinfo{author}{Serna, P.}, \bibinfo{author}{Ortu\~no, M.} \&
  \bibinfo{author}{Somoza, A.~M.}
\newblock \bibinfo{title}{{Deconfined Quantum Criticality, Scaling Violations,
  and Classical Loop Models}}.
\newblock \emph{\bibinfo{journal}{Phys. Rev. X}} \textbf{\bibinfo{volume}{5}},
  \bibinfo{pages}{041048} (\bibinfo{year}{2015}).

\bibitem{Nahum2015prl}
\bibinfo{author}{Nahum, A.}, \bibinfo{author}{Serna, P.},
  \bibinfo{author}{Chalker, J.~T.}, \bibinfo{author}{Ortu\~no, M.} \&
  \bibinfo{author}{Somoza, A.~M.}
\newblock \bibinfo{title}{{Emergent SO(5) Symmetry at the N\'eel to
  Valence-Bond-Solid Transition}}.
\newblock \emph{\bibinfo{journal}{Phys. Rev. Lett.}}
  \textbf{\bibinfo{volume}{115}}, \bibinfo{pages}{267203}
  (\bibinfo{year}{2015}).

\bibitem{Shao2016}
\bibinfo{author}{Shao, H.}, \bibinfo{author}{Guo, W.} \&
  \bibinfo{author}{Sandvik, A.~W.}
\newblock \bibinfo{title}{{Quantum criticality with two length scales}}.
\newblock \emph{\bibinfo{journal}{Science}} \textbf{\bibinfo{volume}{352}},
  \bibinfo{pages}{213--216} (\bibinfo{year}{2016}).

\bibitem{Ma2018prb}
\bibinfo{author}{Ma, N.} \emph{et~al.}
\newblock \bibinfo{title}{{Dynamical signature of fractionalization at a
  deconfined quantum critical point}}.
\newblock \emph{\bibinfo{journal}{Phys. Rev. B}} \textbf{\bibinfo{volume}{98}},
  \bibinfo{pages}{174421} (\bibinfo{year}{2018}).

\bibitem{Wang2017prx}
\bibinfo{author}{Wang, C.}, \bibinfo{author}{Nahum, A.},
  \bibinfo{author}{Metlitski, M.~A.}, \bibinfo{author}{Xu, C.} \&
  \bibinfo{author}{Senthil, T.}
\newblock \bibinfo{title}{{Deconfined Quantum Critical Points: Symmetries and
  Dualities}}.
\newblock \emph{\bibinfo{journal}{Phys. Rev. X}} \textbf{\bibinfo{volume}{7}},
  \bibinfo{pages}{031051} (\bibinfo{year}{2017}).

\bibitem{Ihrig2019prb}
\bibinfo{author}{Ihrig, B.}, \bibinfo{author}{Zerf, N.},
  \bibinfo{author}{Marquard, P.}, \bibinfo{author}{Herbut, I.~F.} \&
  \bibinfo{author}{Scherer, M.~M.}
\newblock \bibinfo{title}{{Abelian Higgs model at four loops, fixed-point
  collision, and deconfined criticality}}.
\newblock \emph{\bibinfo{journal}{Phys. Rev. B}}
  \textbf{\bibinfo{volume}{100}}, \bibinfo{pages}{134507}
  (\bibinfo{year}{2019}).

\bibitem{Zhao2020prl}
\bibinfo{author}{Zhao, B.}, \bibinfo{author}{Takahashi, J.} \&
  \bibinfo{author}{Sandvik, A.~W.}
\newblock \bibinfo{title}{{Multicritical Deconfined Quantum Criticality and
  Lifshitz Point of a Helical Valence-Bond Phase}}.
\newblock \emph{\bibinfo{journal}{Phys. Rev. Lett.}}
  \textbf{\bibinfo{volume}{125}}, \bibinfo{pages}{257204}
  (\bibinfo{year}{2020}).

\bibitem{Ma20}
\bibinfo{author}{Ma, R.} \& \bibinfo{author}{Wang, C.}
\newblock \bibinfo{title}{{Theory of deconfined pseudocriticality}}.
\newblock \emph{\bibinfo{journal}{Phys. Rev. B}}
  \textbf{\bibinfo{volume}{102}}, \bibinfo{pages}{020407}
  (\bibinfo{year}{2020}).

\bibitem{Nahum20}
\bibinfo{author}{Nahum, A.}
\newblock \bibinfo{title}{{Note on Wess-Zumino-Witten models and
  quasiuniversality in $2+1$ dimensions}}.
\newblock \emph{\bibinfo{journal}{Phys. Rev. B}}
  \textbf{\bibinfo{volume}{102}}, \bibinfo{pages}{201116}
  (\bibinfo{year}{2020}).

\bibitem{He21}
\bibinfo{author}{He, Y.-C.}, \bibinfo{author}{Rong, J.} \& \bibinfo{author}{Su,
  N.}
\newblock \bibinfo{title}{{Non-Wilson-Fisher kinks of $O(N)$ numerical
  bootstrap: from the deconfined phase transition to a putative new family of
  CFTs}}.
\newblock \emph{\bibinfo{journal}{SciPost Phys.}}
  \textbf{\bibinfo{volume}{10}}, \bibinfo{pages}{115} (\bibinfo{year}{2021}).

\bibitem{Wang21}
\bibinfo{author}{Wang, Z.}, \bibinfo{author}{Zaletel, M.~P.},
  \bibinfo{author}{Mong, R. S.~K.} \& \bibinfo{author}{Assaad, F.~F.}
\newblock \bibinfo{title}{{Phases of the ($2+1$) Dimensional SO(5) Nonlinear
  Sigma Model with Topological Term}}.
\newblock \emph{\bibinfo{journal}{Phys. Rev. Lett.}}
  \textbf{\bibinfo{volume}{126}}, \bibinfo{pages}{045701}
  (\bibinfo{year}{2021}).

\bibitem{Lu21}
\bibinfo{author}{Lu, D.-C.}, \bibinfo{author}{Xu, C.} \& \bibinfo{author}{You,
  Y.-Z.}
\newblock \bibinfo{title}{{Self-duality protected multicriticality in
  deconfined quantum phase transitions}}.
\newblock \emph{\bibinfo{journal}{Phys. Rev. B}}
  \textbf{\bibinfo{volume}{104}}, \bibinfo{pages}{205142}
  (\bibinfo{year}{2021}).

\bibitem{Yuan23}
\bibinfo{author}{Liao, Y.~D.}, \bibinfo{author}{Pan, G.},
  \bibinfo{author}{Jiang, W.}, \bibinfo{author}{Qi, Y.} \&
  \bibinfo{author}{Meng, Z.~Y.}
\newblock \bibinfo{title}{{The teaching from entanglement: $2$D SU($2$)
  antiferromagnet to valence bond solid deconfined quantum critical points are
  not conformal}}.
\newblock \eprint{Preprint at https://arxiv.org/abs/2302.11742 (2023)}.

\bibitem{Christos2023pnas}
\bibinfo{author}{Christos, M.} \emph{et~al.}
\newblock \bibinfo{title}{{A model of $d$-wave superconductivity,
  antiferromagnetism, and charge order on the square lattice}}.
\newblock \emph{\bibinfo{journal}{Proceedings of the National Academy of
  Sciences}} \textbf{\bibinfo{volume}{120}}, \bibinfo{pages}{(21) e2302701120}
  (\bibinfo{year}{2023}).

\bibitem{Takahashi2020prr}
\bibinfo{author}{Takahashi, J.} \& \bibinfo{author}{Sandvik, A.~W.}
\newblock \bibinfo{title}{{Valence-bond solids, vestigial order, and emergent
  SO(5) symmetry in a two-dimensional quantum magnet}}.
\newblock \emph{\bibinfo{journal}{Phys. Rev. Res.}}
  \textbf{\bibinfo{volume}{2}}, \bibinfo{pages}{033459} (\bibinfo{year}{2020}).

\bibitem{Nakayama2016prl}
\bibinfo{author}{Nakayama, Y.} \& \bibinfo{author}{Ohtsuki, T.}
\newblock \bibinfo{title}{Necessary condition for emergent symmetry from the
  conformal bootstrap}.
\newblock \emph{\bibinfo{journal}{Phys. Rev. Lett.}}
  \textbf{\bibinfo{volume}{117}}, \bibinfo{pages}{131601}
  (\bibinfo{year}{2016}).

\bibitem{Jiang2008jst}
\bibinfo{author}{Jiang, F.-J.}, \bibinfo{author}{Nyfeler, M.},
  \bibinfo{author}{Chandrasekharan, S.} \& \bibinfo{author}{Wiese, U.-J.}
\newblock \bibinfo{title}{From an antiferromagnet to a valence bond solid:
  evidence for a first-order phase transition}.
\newblock \emph{\bibinfo{journal}{Journal of Statistical Mechanics: Theory and
  Experiment}} \textbf{\bibinfo{volume}{2008}}, \bibinfo{pages}{P02009}
  (\bibinfo{year}{2008}).

\bibitem{Li2022jhep}
\bibinfo{author}{Li, Z.}
\newblock \bibinfo{title}{{Bootstrapping conformal QED3 and deconfined quantum
  critical point}}.
\newblock \emph{\bibinfo{journal}{Journal of High Energy Physics}}
  \textbf{\bibinfo{volume}{2022}}, \bibinfo{pages}{5} (\bibinfo{year}{2022}).

\bibitem{DEmidio2023sp}
\bibinfo{author}{D'Emidio, J.}, \bibinfo{author}{Eberharter, A.~A.} \&
  \bibinfo{author}{L\"auchli, A.~M.}
\newblock \bibinfo{title}{{Diagnosing weakly first-order phase transitions by
  coupling to order parameters}}.
\newblock \emph{\bibinfo{journal}{SciPost Phys.}}
  \textbf{\bibinfo{volume}{15}}, \bibinfo{pages}{061} (\bibinfo{year}{2023}).

\bibitem{Chen2024prl}
\bibinfo{author}{Chen, B.-B.}, \bibinfo{author}{Zhang, X.},
  \bibinfo{author}{Wang, Y.}, \bibinfo{author}{Sun, K.} \&
  \bibinfo{author}{Meng, Z.~Y.}
\newblock \bibinfo{title}{{Phases of $(2+1)\mathrm{D}$ SO(5) Nonlinear Sigma
  Model with a Topological Term on a Sphere: Multicritical Point and Disorder
  Phase}}.
\newblock \emph{\bibinfo{journal}{Phys. Rev. Lett.}}
  \textbf{\bibinfo{volume}{132}}, \bibinfo{pages}{246503}
  (\bibinfo{year}{2024}).

\bibitem{Takahashi2024}
\bibinfo{author}{Takahashi, J.}, \bibinfo{author}{Shao, H.},
  \bibinfo{author}{Zhao, B.}, \bibinfo{author}{Guo, W.} \&
  \bibinfo{author}{Sandvik, A.~W.}
\newblock \bibinfo{title}{{SO($5$) multicriticality in two-dimensional quantum
  magnets}}.
\newblock \eprint{Preprint at https://arxiv.org/abs/2405.06607 (2024)}.

\bibitem{Deng2024prl}
\bibinfo{author}{Deng, Z.}, \bibinfo{author}{Liu, L.}, \bibinfo{author}{Guo,
  W.} \& \bibinfo{author}{Lin, H.-Q.}
\newblock \bibinfo{title}{{Diagnosing Quantum Phase Transition Order and
  Deconfined Criticality via Entanglement Entropy}}.
\newblock \emph{\bibinfo{journal}{Phys. Rev. Lett.}}
  \textbf{\bibinfo{volume}{133}}, \bibinfo{pages}{100402}
  (\bibinfo{year}{2024}).

\bibitem{DEmidio2024prl}
\bibinfo{author}{D'Emidio, J.} \& \bibinfo{author}{Sandvik, A.~W.}
\newblock \bibinfo{title}{{Entanglement Entropy and Deconfined Criticality:
  Emergent SO($5$) Symmetry and Proper Lattice Bipartition}}.
\newblock \emph{\bibinfo{journal}{Phys. Rev. Lett.}}
  \textbf{\bibinfo{volume}{133}}, \bibinfo{pages}{166702}
  (\bibinfo{year}{2024}).

\bibitem{Chester2024prl}
\bibinfo{author}{Chester, S.~M.} \& \bibinfo{author}{Su, N.}
\newblock \bibinfo{title}{{Bootstrapping Deconfined Quantum Tricriticality}}.
\newblock \emph{\bibinfo{journal}{Phys. Rev. Lett.}}
  \textbf{\bibinfo{volume}{132}}, \bibinfo{pages}{111601}
  (\bibinfo{year}{2024}).

\bibitem{Zhou2024prx}
\bibinfo{author}{Zhou, Z.}, \bibinfo{author}{Hu, L.}, \bibinfo{author}{Zhu, W.}
  \& \bibinfo{author}{He, Y.-C.}
\newblock \bibinfo{title}{{SO($5$) Deconfined Phase Transition under the
  Fuzzy-Sphere Microscope: Approximate Conformal Symmetry, Pseudo-Criticality,
  and Operator Spectrum}}.
\newblock \emph{\bibinfo{journal}{Phys. Rev. X}} \textbf{\bibinfo{volume}{14}},
  \bibinfo{pages}{021044} (\bibinfo{year}{2024}).

\bibitem{Chen2024prb}
\bibinfo{author}{Chen, B.-B.}, \bibinfo{author}{Zhang, X.} \&
  \bibinfo{author}{Meng, Z.~Y.}
\newblock \bibinfo{title}{{Emergent conformal symmetry at the multicritical
  point of $(2+1)\mathrm{D}$ SO(5) model with Wess-Zumino-Witten term on a
  sphere}}.
\newblock \emph{\bibinfo{journal}{Phys. Rev. B}}
  \textbf{\bibinfo{volume}{110}}, \bibinfo{pages}{125153}
  (\bibinfo{year}{2024}).

\bibitem{Kibble1976jpa}
\bibinfo{author}{Kibble, T. W.~B.}
\newblock \bibinfo{title}{{Topology of cosmic domains and strings}}.
\newblock \emph{\bibinfo{journal}{Journal of Physics A: Mathematical and
  General}} \textbf{\bibinfo{volume}{9}}, \bibinfo{pages}{1387}
  (\bibinfo{year}{1976}).

\bibitem{Zurek1985nat}
\bibinfo{author}{Zurek, W.~H.}
\newblock \bibinfo{title}{Cosmological experiments in superfluid helium?}
\newblock \emph{\bibinfo{journal}{Nature}} \textbf{\bibinfo{volume}{317}},
  \bibinfo{pages}{505--508} (\bibinfo{year}{1985}).

\bibitem{Polkovnikov2011rmp}
\bibinfo{author}{Polkovnikov, A.}, \bibinfo{author}{Sengupta, K.},
  \bibinfo{author}{Silva, A.} \& \bibinfo{author}{Vengalattore, M.}
\newblock \bibinfo{title}{{Colloquium: Nonequilibrium dynamics of closed
  interacting quantum systems}}.
\newblock \emph{\bibinfo{journal}{Rev. Mod. Phys.}}
  \textbf{\bibinfo{volume}{83}}, \bibinfo{pages}{863--883}
  (\bibinfo{year}{2011}).

\bibitem{Dziarmaga2010}
\bibinfo{author}{Dziarmaga, J.}
\newblock \bibinfo{title}{{Dynamics of a quantum phase transition and
  relaxation to a steady state}}.
\newblock \emph{\bibinfo{journal}{Advances in Physics}}
  \textbf{\bibinfo{volume}{59}}, \bibinfo{pages}{1063--1189}
  (\bibinfo{year}{2010}).

\bibitem{Zhong2005prb}
\bibinfo{author}{Zhong, F.} \& \bibinfo{author}{Xu, Z.}
\newblock \bibinfo{title}{{Dynamic Monte Carlo renormalization group
  determination of critical exponents with linearly changing temperature}}.
\newblock \emph{\bibinfo{journal}{Phys. Rev. B}} \textbf{\bibinfo{volume}{71}},
  \bibinfo{pages}{132402} (\bibinfo{year}{2005}).

\bibitem{DeGrandi2011prb}
\bibinfo{author}{De~Grandi, C.}, \bibinfo{author}{Polkovnikov, A.} \&
  \bibinfo{author}{Sandvik, A.~W.}
\newblock \bibinfo{title}{{Universal nonequilibrium quantum dynamics in
  imaginary time}}.
\newblock \emph{\bibinfo{journal}{Phys. Rev. B}} \textbf{\bibinfo{volume}{84}},
  \bibinfo{pages}{224303} (\bibinfo{year}{2011}).

\bibitem{Liu2014prb}
\bibinfo{author}{Liu, C.-W.}, \bibinfo{author}{Polkovnikov, A.} \&
  \bibinfo{author}{Sandvik, A.~W.}
\newblock \bibinfo{title}{{Dynamic scaling at classical phase transitions
  approached through nonequilibrium quenching}}.
\newblock \emph{\bibinfo{journal}{Phys. Rev. B}} \textbf{\bibinfo{volume}{89}},
  \bibinfo{pages}{054307} (\bibinfo{year}{2014}).

\bibitem{Huang2014prb}
\bibinfo{author}{Huang, Y.}, \bibinfo{author}{Yin, S.}, \bibinfo{author}{Feng,
  B.} \& \bibinfo{author}{Zhong, F.}
\newblock \bibinfo{title}{{Kibble-Zurek mechanism and finite-time scaling}}.
\newblock \emph{\bibinfo{journal}{Phys. Rev. B}} \textbf{\bibinfo{volume}{90}},
  \bibinfo{pages}{134108} (\bibinfo{year}{2014}).

\bibitem{Lin2014natphy}
\bibinfo{author}{Lin, S.-Z.}, \bibinfo{author}{Wang, X.},
  \bibinfo{author}{Kamiya, Y.} \& \bibinfo{author}{{\it et al}}.
\newblock \bibinfo{title}{{Topological defects as relics of emergent continuous
  symmetry and Higgs condensation of disorder in ferroelectrics}}.
\newblock \emph{\bibinfo{journal}{Nature Physics}}
  \textbf{\bibinfo{volume}{10}}, \bibinfo{pages}{970--977}
  (\bibinfo{year}{2014}).

\bibitem{Clark2016sci}
\bibinfo{author}{Clark, L.~W.}, \bibinfo{author}{Feng, L.} \&
  \bibinfo{author}{Chin, C.}
\newblock \bibinfo{title}{{Universal space-time scaling symmetry in the
  dynamics of bosons across a quantum phase transition}}.
\newblock \emph{\bibinfo{journal}{Science}} \textbf{\bibinfo{volume}{354}},
  \bibinfo{pages}{606--610} (\bibinfo{year}{2016}).

\bibitem{Rysti2021prl}
\bibinfo{author}{Rysti, J.} \emph{et~al.}
\newblock \bibinfo{title}{{Suppressing the Kibble-Zurek Mechanism by a
  Symmetry-Violating Bias}}.
\newblock \emph{\bibinfo{journal}{Phys. Rev. Lett.}}
  \textbf{\bibinfo{volume}{127}}, \bibinfo{pages}{115702}
  (\bibinfo{year}{2021}).

\bibitem{Rams2019prl}
\bibinfo{author}{Rams, M.~M.}, \bibinfo{author}{Dziarmaga, J.} \&
  \bibinfo{author}{Zurek, W.~H.}
\newblock \bibinfo{title}{{Symmetry Breaking Bias and the Dynamics of a Quantum
  Phase Transition}}.
\newblock \emph{\bibinfo{journal}{Phys. Rev. Lett.}}
  \textbf{\bibinfo{volume}{123}}, \bibinfo{pages}{130603}
  (\bibinfo{year}{2019}).

\bibitem{Keesling2019nat}
\bibinfo{author}{Keesling, A.}, \bibinfo{author}{Omran, A.},
  \bibinfo{author}{Levine, H.} \& \bibinfo{author}{{\it et al}}.
\newblock \bibinfo{title}{{Quantum Kibble–Zurek mechanism and critical
  dynamics on a programmable Rydberg simulator}}.
\newblock \emph{\bibinfo{journal}{Nature}} \textbf{\bibinfo{volume}{568}},
  \bibinfo{pages}{207--211} (\bibinfo{year}{2019}).

\bibitem{King2023nat}
\bibinfo{author}{King, A.~D.} \emph{et~al.}
\newblock \bibinfo{title}{Quantum critical dynamics in a 5,000-qubit
  programmable spin glass}.
\newblock \emph{\bibinfo{journal}{Nature}} \textbf{\bibinfo{volume}{617}},
  \bibinfo{pages}{61--66} (\bibinfo{year}{2023}).

\bibitem{Suzuki2024prl}
\bibinfo{author}{Suzuki, F.} \& \bibinfo{author}{Zurek, W.~H.}
\newblock \bibinfo{title}{Topological defect formation in a phase transition
  with tunable order}.
\newblock \emph{\bibinfo{journal}{Phys. Rev. Lett.}}
  \textbf{\bibinfo{volume}{132}}, \bibinfo{pages}{241601}
  (\bibinfo{year}{2024}).

\bibitem{Gong2010njp}
\bibinfo{author}{Gong, S.}, \bibinfo{author}{Zhong, F.},
  \bibinfo{author}{Huang, X.} \& \bibinfo{author}{Fan, S.}
\newblock \bibinfo{title}{{Finite-time scaling via linear driving}}.
\newblock \emph{\bibinfo{journal}{New Journal of Physics}}
  \textbf{\bibinfo{volume}{12}}, \bibinfo{pages}{043036}
  (\bibinfo{year}{2010}).

\bibitem{Liu2015pre}
\bibinfo{author}{Liu, C.-W.}, \bibinfo{author}{Polkovnikov, A.},
  \bibinfo{author}{Sandvik, A.~W.} \& \bibinfo{author}{Young, A.~P.}
\newblock \bibinfo{title}{Universal dynamic scaling in three-dimensional ising
  spin glasses}.
\newblock \emph{\bibinfo{journal}{Phys. Rev. E}} \textbf{\bibinfo{volume}{92}},
  \bibinfo{pages}{022128} (\bibinfo{year}{2015}).

\bibitem{Zurek2005prl}
\bibinfo{author}{Zurek, W.~H.}, \bibinfo{author}{Dorner, U.} \&
  \bibinfo{author}{Zoller, P.}
\newblock \bibinfo{title}{{Dynamics of a Quantum Phase Transition}}.
\newblock \emph{\bibinfo{journal}{Phys. Rev. Lett.}}
  \textbf{\bibinfo{volume}{95}}, \bibinfo{pages}{105701}
  (\bibinfo{year}{2005}).

\bibitem{Dziarmaga2005prl}
\bibinfo{author}{Dziarmaga, J.}
\newblock \bibinfo{title}{{Dynamics of a Quantum Phase Transition: Exact
  Solution of the Quantum Ising Model}}.
\newblock \emph{\bibinfo{journal}{Phys. Rev. Lett.}}
  \textbf{\bibinfo{volume}{95}}, \bibinfo{pages}{245701}
  (\bibinfo{year}{2005}).

\bibitem{Polkovnikov2005prb}
\bibinfo{author}{Polkovnikov, A.}
\newblock \bibinfo{title}{{Universal adiabatic dynamics in the vicinity of a
  quantum critical point}}.
\newblock \emph{\bibinfo{journal}{Phys. Rev. B}} \textbf{\bibinfo{volume}{72}},
  \bibinfo{pages}{161201} (\bibinfo{year}{2005}).

\bibitem{Schmitt2022sciadv}
\bibinfo{author}{Schmitt, M.}, \bibinfo{author}{Rams, M.~M.},
  \bibinfo{author}{Dziarmaga, J.}, \bibinfo{author}{Heyl, M.} \&
  \bibinfo{author}{Zurek, W.~H.}
\newblock \bibinfo{title}{{Quantum phase transition dynamics in the
  two-dimensional transverse-field Ising model}}.
\newblock \emph{\bibinfo{journal}{Science Advances}}
  \textbf{\bibinfo{volume}{8}}, \bibinfo{pages}{eabl6850}
  (\bibinfo{year}{2022}).

\bibitem{Shu2022prl}
\bibinfo{author}{Shu, Y.-R.}, \bibinfo{author}{Jian, S.-K.} \&
  \bibinfo{author}{Yin, S.}
\newblock \bibinfo{title}{{Nonequilibrium Dynamics of Deconfined Quantum
  Critical Point in Imaginary Time}}.
\newblock \emph{\bibinfo{journal}{Phys. Rev. Lett.}}
  \textbf{\bibinfo{volume}{128}}, \bibinfo{pages}{020601}
  (\bibinfo{year}{2022}).

\bibitem{Shao2020prl}
\bibinfo{author}{Shao, H.}, \bibinfo{author}{Guo, W.} \&
  \bibinfo{author}{Sandvik, A.~W.}
\newblock \bibinfo{title}{{Monte Carlo Renormalization Flows in the Space of
  Relevant and Irrelevant Operators: Application to Three-Dimensional Clock
  Models}}.
\newblock \emph{\bibinfo{journal}{Phys. Rev. Lett.}}
  \textbf{\bibinfo{volume}{124}}, \bibinfo{pages}{080602}
  (\bibinfo{year}{2020}).

\bibitem{Patil2021prb}
\bibinfo{author}{Patil, P.}, \bibinfo{author}{Shao, H.} \&
  \bibinfo{author}{Sandvik, A.~W.}
\newblock \bibinfo{title}{{Unconventional U(1) to ${Z}_{q}$ crossover in
  quantum and classical $q$-state clock models}}.
\newblock \emph{\bibinfo{journal}{Phys. Rev. B}}
  \textbf{\bibinfo{volume}{103}}, \bibinfo{pages}{054418}
  (\bibinfo{year}{2021}).

\bibitem{Zhao2020cpl}
\bibinfo{author}{Sandvik, A.~W.} \& \bibinfo{author}{Zhao, B.}
\newblock \bibinfo{title}{{Consistent Scaling Exponents at the Deconfined
  Quantum-Critical Point}}.
\newblock \emph{\bibinfo{journal}{Chinese Physics Letters}}
  \textbf{\bibinfo{volume}{37}}, \bibinfo{pages}{057502}
  (\bibinfo{year}{2020}).

\bibitem{Tang2013prl}
\bibinfo{author}{Tang, Y.} \& \bibinfo{author}{Sandvik, A.~W.}
\newblock \bibinfo{title}{{Confinement and Deconfinement of Spinons in Two
  Dimensions}}.
\newblock \emph{\bibinfo{journal}{Phys. Rev. Lett.}}
  \textbf{\bibinfo{volume}{110}}, \bibinfo{pages}{217213}
  (\bibinfo{year}{2013}).

\bibitem{Oshikawa2000prb}
\bibinfo{author}{Oshikawa, M.}
\newblock \bibinfo{title}{{Ordered phase and scaling in ${Z}_{n}$ models and
  the three-state antiferromagnetic Potts model in three dimensions}}.
\newblock \emph{\bibinfo{journal}{Phys. Rev. B}} \textbf{\bibinfo{volume}{61}},
  \bibinfo{pages}{3430--3434} (\bibinfo{year}{2000}).

\bibitem{Chester2020jhep}
\bibinfo{author}{Chester, S.~M.} \emph{et~al.}
\newblock \bibinfo{title}{{Carving out OPE space and precise $O(2)$ model
  critical exponents}}.
\newblock \emph{\bibinfo{journal}{Journal of High Energy Physics}}
  \textbf{\bibinfo{volume}{2020}}, \bibinfo{pages}{142} (\bibinfo{year}{2020}).

\bibitem{Adzhemyan2022pa}
\bibinfo{author}{Adzhemyan, L.} \emph{et~al.}
\newblock \bibinfo{title}{{Model A of critical dynamics: 5-loop $\varepsilon$
  expansion study}}.
\newblock \emph{\bibinfo{journal}{Physica A: Statistical Mechanics and its
  Applications}} \textbf{\bibinfo{volume}{600}}, \bibinfo{pages}{127530}
  (\bibinfo{year}{2022}).

\bibitem{Campostrini2001prb}
\bibinfo{author}{Campostrini, M.}, \bibinfo{author}{Hasenbusch, M.},
  \bibinfo{author}{Pelissetto, A.}, \bibinfo{author}{Rossi, P.} \&
  \bibinfo{author}{Vicari, E.}
\newblock \bibinfo{title}{{Critical behavior of the three-dimensional
  $\mathrm{XY}$ universality class}}.
\newblock \emph{\bibinfo{journal}{Phys. Rev. B}} \textbf{\bibinfo{volume}{63}},
  \bibinfo{pages}{214503} (\bibinfo{year}{2001}).

\bibitem{Campostrini2006prb}
\bibinfo{author}{Campostrini, M.}, \bibinfo{author}{Hasenbusch, M.},
  \bibinfo{author}{Pelissetto, A.} \& \bibinfo{author}{Vicari, E.}
\newblock \bibinfo{title}{{Theoretical estimates of the critical exponents of
  the superfluid transition in $^{4}\mathrm{He}$ by lattice methods}}.
\newblock \emph{\bibinfo{journal}{Phys. Rev. B}} \textbf{\bibinfo{volume}{74}},
  \bibinfo{pages}{144506} (\bibinfo{year}{2006}).

\bibitem{DeGrandi2013jpcm}
\bibinfo{author}{De~Grandi, C.}, \bibinfo{author}{Polkovnikov, A.} \&
  \bibinfo{author}{Sandvik, A.~W.}
\newblock \bibinfo{title}{{Microscopic theory of non-adiabatic response in real
  and imaginary time}}.
\newblock \emph{\bibinfo{journal}{Journal of Physics: Condensed Matter}}
  \textbf{\bibinfo{volume}{25}}, \bibinfo{pages}{404216}
  (\bibinfo{year}{2013}).

\bibitem{Liu2013prb}
\bibinfo{author}{Liu, C.-W.}, \bibinfo{author}{Polkovnikov, A.} \&
  \bibinfo{author}{Sandvik, A.~W.}
\newblock \bibinfo{title}{{Quasi-adiabatic quantum Monte Carlo algorithm for
  quantum evolution in imaginary time}}.
\newblock \emph{\bibinfo{journal}{Phys. Rev. B}} \textbf{\bibinfo{volume}{87}},
  \bibinfo{pages}{174302} (\bibinfo{year}{2013}).

\bibitem{Sandvik2010review}
\bibinfo{author}{Sandvik, A.~W.}
\newblock \bibinfo{title}{{Computational Studies of Quantum Spin Systems}}.
\newblock \emph{\bibinfo{journal}{AIP Conference Proceedings}}
  \textbf{\bibinfo{volume}{1297}}, \bibinfo{pages}{135--338}
  (\bibinfo{year}{2010}).

\bibitem{Tang2011prl}
\bibinfo{author}{Tang, Y.} \& \bibinfo{author}{Sandvik, A.~W.}
\newblock \bibinfo{title}{{Method to Characterize Spinons as Emergent
  Elementary Particles}}.
\newblock \emph{\bibinfo{journal}{Phys. Rev. Lett.}}
  \textbf{\bibinfo{volume}{107}}, \bibinfo{pages}{157201}
  (\bibinfo{year}{2011}).

\bibitem{Beach2006npb}
\bibinfo{author}{Beach, K.} \& \bibinfo{author}{Sandvik, A.~W.}
\newblock \bibinfo{title}{{Some formal results for the valence bond basis}}.
\newblock \emph{\bibinfo{journal}{Nuclear Physics B}}
  \textbf{\bibinfo{volume}{750}}, \bibinfo{pages}{142--178}
  (\bibinfo{year}{2006}).

\bibitem{Suzuki1976}
\bibinfo{author}{Suzuki, M.}
\newblock \bibinfo{title}{{Relationship between d-Dimensional Quantal Spin
  Systems and (d+1)-Dimensional Ising Systems: Equivalence, Critical Exponents
  and Systematic Approximants of the Partition Function and Spin Correlations
  }}.
\newblock \emph{\bibinfo{journal}{Prog. Theor. Phys.}}
  \textbf{\bibinfo{volume}{56}}, \bibinfo{pages}{1454} (\bibinfo{year}{1976}).

\bibitem{Sachdev1999}
\bibinfo{author}{Sachdev, S.}
\newblock \emph{\bibinfo{title}{{Quantum Phase Transitions}}}
  (\bibinfo{publisher}{Cambridge University Press}, \bibinfo{year}{1999}).

\bibitem{Sondhi1997rmp}
\bibinfo{author}{Sondhi, S.~L.}, \bibinfo{author}{Girvin, S.~M.},
  \bibinfo{author}{Carini, J.~P.} \& \bibinfo{author}{Shahar, D.}
\newblock \bibinfo{title}{{Continuous quantum phase transitions}}.
\newblock \emph{\bibinfo{journal}{Rev. Mod. Phys.}}
  \textbf{\bibinfo{volume}{69}}, \bibinfo{pages}{315--333}
  (\bibinfo{year}{1997}).

\bibitem{Vojta2003rpp}
\bibinfo{author}{Vojta, M.}
\newblock \bibinfo{title}{{Quantum phase transitions}}.
\newblock \emph{\bibinfo{journal}{Rep. Prog. Phys.}}
  \textbf{\bibinfo{volume}{66}}, \bibinfo{pages}{2069} (\bibinfo{year}{2003}).

\bibitem{Shu2017prb}
\bibinfo{author}{Shu, Y.-R.}, \bibinfo{author}{Yin, S.} \&
  \bibinfo{author}{Yao, D.-X.}
\newblock \bibinfo{title}{{Universal short-time quantum critical dynamics of
  finite-size systems}}.
\newblock \emph{\bibinfo{journal}{Phys. Rev. B}} \textbf{\bibinfo{volume}{96}},
  \bibinfo{pages}{094304} (\bibinfo{year}{2017}).

\bibitem{Hasenbusch2010prb}
\bibinfo{author}{Hasenbusch, M.}
\newblock \bibinfo{title}{{Finite size scaling study of lattice models in the
  three-dimensional Ising universality class}}.
\newblock \emph{\bibinfo{journal}{Phys. Rev. B}} \textbf{\bibinfo{volume}{82}},
  \bibinfo{pages}{174433} (\bibinfo{year}{2010}).

\bibitem{Zhong2011book}
\bibinfo{author}{Zhong, F.}
\newblock \bibinfo{title}{{Finite-time Scaling and its Applications to
  Continuous Phase Transitions}}.
\newblock In \bibinfo{editor}{Mordechai, S.} (ed.)
  \emph{\bibinfo{booktitle}{Applications of Monte Carlo Method in Science and
  Engineering}}, chap.~\bibinfo{chapter}{18} (\bibinfo{publisher}{IntechOpen},
  \bibinfo{address}{Rijeka}, \bibinfo{year}{2011}).

\bibitem{Jose1977prb}
\bibinfo{author}{Jos\'e, J.~V.}, \bibinfo{author}{Kadanoff, L.~P.},
  \bibinfo{author}{Kirkpatrick, S.} \& \bibinfo{author}{Nelson, D.~R.}
\newblock \bibinfo{title}{{Renormalization, vortices, and symmetry-breaking
  perturbations in the two-dimensional planar model}}.
\newblock \emph{\bibinfo{journal}{Phys. Rev. B}} \textbf{\bibinfo{volume}{16}},
  \bibinfo{pages}{1217--1241} (\bibinfo{year}{1977}).

\bibitem{Pujari2015prb}
\bibinfo{author}{Pujari, S.}, \bibinfo{author}{Alet, F.} \&
  \bibinfo{author}{Damle, K.}
\newblock \bibinfo{title}{{Transitions to valence-bond solid order in a
  honeycomb lattice antiferromagnet}}.
\newblock \emph{\bibinfo{journal}{Phys. Rev. B}} \textbf{\bibinfo{volume}{91}},
  \bibinfo{pages}{104411} (\bibinfo{year}{2015}).

\bibitem{Lou2007prl}
\bibinfo{author}{Lou, J.}, \bibinfo{author}{Sandvik, A.~W.} \&
  \bibinfo{author}{Balents, L.}
\newblock \bibinfo{title}{{Emergence of U(1) Symmetry in the 3D $XY$ Model with
  ${Z}_{q}$ Anisotropy}}.
\newblock \emph{\bibinfo{journal}{Phys. Rev. Lett.}}
  \textbf{\bibinfo{volume}{99}}, \bibinfo{pages}{207203}
  (\bibinfo{year}{2007}).

\bibitem{Okubo2015prb}
\bibinfo{author}{Okubo, T.}, \bibinfo{author}{Oshikawa, K.},
  \bibinfo{author}{Watanabe, H.} \& \bibinfo{author}{Kawashima, N.}
\newblock \bibinfo{title}{{Scaling relation for dangerously irrelevant
  symmetry-breaking fields}}.
\newblock \emph{\bibinfo{journal}{Phys. Rev. B}} \textbf{\bibinfo{volume}{91}},
  \bibinfo{pages}{174417} (\bibinfo{year}{2015}).

\bibitem{Leonard2015prl}
\bibinfo{author}{L\'eonard, F.} \& \bibinfo{author}{Delamotte, B.}
\newblock \bibinfo{title}{{Critical Exponents Can Be Different on the Two Sides
  of a Transition: A Generic Mechanism}}.
\newblock \emph{\bibinfo{journal}{Phys. Rev. Lett.}}
  \textbf{\bibinfo{volume}{115}}, \bibinfo{pages}{200601}
  (\bibinfo{year}{2015}).

\bibitem{Hove2003pre}
\bibinfo{author}{Hove, J.} \& \bibinfo{author}{Sudb\o{}, A.}
\newblock \bibinfo{title}{{Criticality versus q in the $(2+1)$-dimensional
  ${Z}_{q}$ clock model}}.
\newblock \emph{\bibinfo{journal}{Phys. Rev. E}} \textbf{\bibinfo{volume}{68}},
  \bibinfo{pages}{046107} (\bibinfo{year}{2003}).

\bibitem{Hasenbusch2011prb}
\bibinfo{author}{Hasenbusch, M.} \& \bibinfo{author}{Vicari, E.}
\newblock \bibinfo{title}{{Anisotropic perturbations in three-dimensional
  O($N$)-symmetric vector models}}.
\newblock \emph{\bibinfo{journal}{Phys. Rev. B}} \textbf{\bibinfo{volume}{84}},
  \bibinfo{pages}{125136} (\bibinfo{year}{2011}).

\bibitem{Chlebicki2022pre}
\bibinfo{author}{Chlebicki, A.}, \bibinfo{author}{S\'anchez-Villalobos, C.~A.},
  \bibinfo{author}{Jakubczyk, P.} \& \bibinfo{author}{Wschebor, N.}
\newblock \bibinfo{title}{{${\mathbb{Z}}_{4}$-symmetric perturbations to the
  $XY$ model from functional renormalization}}.
\newblock \emph{\bibinfo{journal}{Phys. Rev. E}}
  \textbf{\bibinfo{volume}{106}}, \bibinfo{pages}{064135}
  (\bibinfo{year}{2022}).

\bibitem{Ueno1991prb}
\bibinfo{author}{Ueno, Y.} \& \bibinfo{author}{Mitsubo, K.}
\newblock \bibinfo{title}{{Incompletely ordered phase in the three-dimensional
  six-state clock model: Evidence for an absence of ordered phases of XY
  character}}.
\newblock \emph{\bibinfo{journal}{Phys. Rev. B}} \textbf{\bibinfo{volume}{43}},
  \bibinfo{pages}{8654--8657} (\bibinfo{year}{1991}).

\bibitem{Chubukov1994prb}
\bibinfo{author}{Chubukov, A.~V.}, \bibinfo{author}{Sachdev, S.} \&
  \bibinfo{author}{Ye, J.}
\newblock \bibinfo{title}{Theory of two-dimensional quantum heisenberg
  antiferromagnets with a nearly critical ground state}.
\newblock \emph{\bibinfo{journal}{Phys. Rev. B}} \textbf{\bibinfo{volume}{49}},
  \bibinfo{pages}{11919--11961} (\bibinfo{year}{1994}).

\bibitem{Miyashita1997jpsj}
\bibinfo{author}{Miyashita, S.}
\newblock \bibinfo{title}{{Nature of the Ordered Phase and the Critical
  Properties of the Three Dimensional Six-State Clock Model}}.
\newblock \emph{\bibinfo{journal}{Journal of the Physical Society of Japan}}
  \textbf{\bibinfo{volume}{66}}, \bibinfo{pages}{3411--3420}
  (\bibinfo{year}{1997}).

\bibitem{Zhitomirsky2014prb}
\bibinfo{author}{Zhitomirsky, M.~E.}, \bibinfo{author}{Holdsworth, P. C.~W.} \&
  \bibinfo{author}{Moessner, R.}
\newblock \bibinfo{title}{{Nature of finite-temperature transition in
  anisotropic pyrochlore
  ${\mathrm{Er}}_{2}{\mathrm{Ti}}_{2}{\mathrm{O}}_{7}$}}.
\newblock \emph{\bibinfo{journal}{Phys. Rev. B}} \textbf{\bibinfo{volume}{89}},
  \bibinfo{pages}{140403} (\bibinfo{year}{2014}).

\bibitem{Banerjee2018prl}
\bibinfo{author}{Banerjee, D.}, \bibinfo{author}{Chandrasekharan, S.} \&
  \bibinfo{author}{Orlando, D.}
\newblock \bibinfo{title}{Conformal dimensions via large charge expansion}.
\newblock \emph{\bibinfo{journal}{Phys. Rev. Lett.}}
  \textbf{\bibinfo{volume}{120}}, \bibinfo{pages}{061603}
  (\bibinfo{year}{2018}).

\bibitem{Hasenbusch2019prb}
\bibinfo{author}{Hasenbusch, M.}
\newblock \bibinfo{title}{{Monte Carlo study of an improved clock model in
  three dimensions}}.
\newblock \emph{\bibinfo{journal}{Phys. Rev. B}}
  \textbf{\bibinfo{volume}{100}}, \bibinfo{pages}{224517}
  (\bibinfo{year}{2019}).

\end{thebibliography}

\end{document}